\documentclass[aps,prb,twocolumn,notitlepage,superscriptaddress,longbibliography,preprintnumbers]{revtex4-2}
\usepackage{times}
\usepackage{graphicx}
\usepackage[whole]{bxcjkjatype} 

\usepackage[top=30truemm,bottom=30truemm,left=25truemm,right=25truemm]{geometry}

\usepackage[colorlinks=true,linktoc=page,citecolor=red,linkcolor=blue]{hyperref}	
\graphicspath{{./figs/}}  
\usepackage[usenames]{xcolor}
\usepackage{amsmath,amssymb,amsfonts}	
\usepackage{bm,braket,array}
\usepackage{multirow}
\usepackage[all]{xy}
\usepackage{amsthm}
\usepackage{amscd}
\usepackage{slashed} 
\usepackage[normalem]{ulem} 

\theoremstyle{definition}

\theoremstyle{remark}

\def\Tr{{\rm Tr }}
\def\tr{{\rm tr\,}}

\def\p{\partial}

\def\ker{{\rm Ker\,}}

\def\Hom{{\rm Hom}}

\def\ua{\uparrow}
\def\da{\downarrow}

\newcommand{\s}{\sigma}
\newcommand{\g}{\gamma}

\newcommand{\C}{\mathbb{C}}

\newcommand{\R}{\mathbb{R}}
\newcommand{\Z}{\mathbb{Z}}

\newcommand{\bk}{{\bm{k}}}

\def\widebar{\accentset{{\cc@style\underline{\mskip10mu}}}} 
\def\wideubar{\underaccent{{\cc@style\underline{\mskip10mu}}}} 

\begin{document}

\title{On adiabatic cycles of quantum spin systems}
\author{Ken Shiozaki}
\email{ken.shiozaki@yukawa.kyoto-u.ac.jp}
\affiliation{Yukawa Institute for Theoretical Physics, Kyoto University, Kyoto 606-8502, Japan}

\date{\today} 
\preprint{YITP-21-121}

\begin{abstract}
Motivated by the $\Omega$-spectrum proposal of unique gapped ground states by Kitaev~\cite{Kit13}, we study adiabatic cycles in gapped quantum spin systems from various perspectives.
We give a few exactly solvable models in one and two spatial dimensions and discuss how nontrivial adiabatic cycles are detected. 
For one spatial dimension, we study the adiabatic cycle in detail with the matrix product state and show that the symmetry charge can act on the space of matrices without changing the physical states, which leads to nontrivial loops with symmetry charges. 
For generic spatial dimensions, based on the Bockstein isomorphism $H^d(G,U(1)) \cong H^{d+1}(G,\Z)$, we study a group cohomology model of the adiabatic cycle that pumps a symmetry-protected topological phase on the boundary by one period. 
It is shown that the spatial texture of the adiabatic Hamiltonian traps a symmetry-protected topological phase in one dimension lower. 
\end{abstract}

\maketitle



\section{Introduction}
In the last fifteen years, understanding the phase structure of the gapped ground state of quantum many-body systems has been progressed. 
An equivalence class of gapped ground states by identifying each other without a phase transition is called a topological phase. 
In particular, the topological phases with no ground state degeneracy for any closed space manifolds are called invertible phases or symmetry-protected topological (SPT) phases.
Invertible phases have been studied from various points of view, including invertible phases in free fermions~\cite{SRFL08,Kit09}, classification and model construction in quantum spin systems using group cohomology~\cite{PETO12, CGW11, SPC11, CGLW13, LG12}, and classification of topological response actions using cobordism groups~\cite{Kap14, KTTW15, FH16}.

This paper is motivated by the Kitaev proposal that invertible states form an $\Omega$-spectrum in generalized cohomology theory~\cite{Kit11,Kit13,Kit15}.
Let $E_d$ be the ``space of invertible states" in $d$ spatial dimension, which has not yet been rigorously defined. 
The space $E_d$ is equipped with a basepoint as the trivial tensor product state $\ket{0}$. 
The sequence of spaces $\{E_d\}_{d \in \Z}$ is called an $\Omega$-spectrum if and only if the based loop space $\Omega E_{d+1} = \{\ell: S^1 \to E_{d+1} | \ell(0)=\ell(1) = \ket{0}\}$, the space of loops in $(d+1)$-dimensional invertible states that start and end at the trivial state, is homotopically equivalent to $E_d$, the space of invertible states one dimension lower. 
Mathematically, an $\Omega$-spectrum defines a generalized cohomology theory. 
Thus, it is predicted that a generalized cohomology theory gives the classification of invertible phases. See \cite{Xio18} for a review of this perspective for lattice models, and \cite{GT17} for field theories. 

\begin{figure}[!]
\centering
\includegraphics[width=\linewidth, trim=0cm 0cm 0cm 0cm]{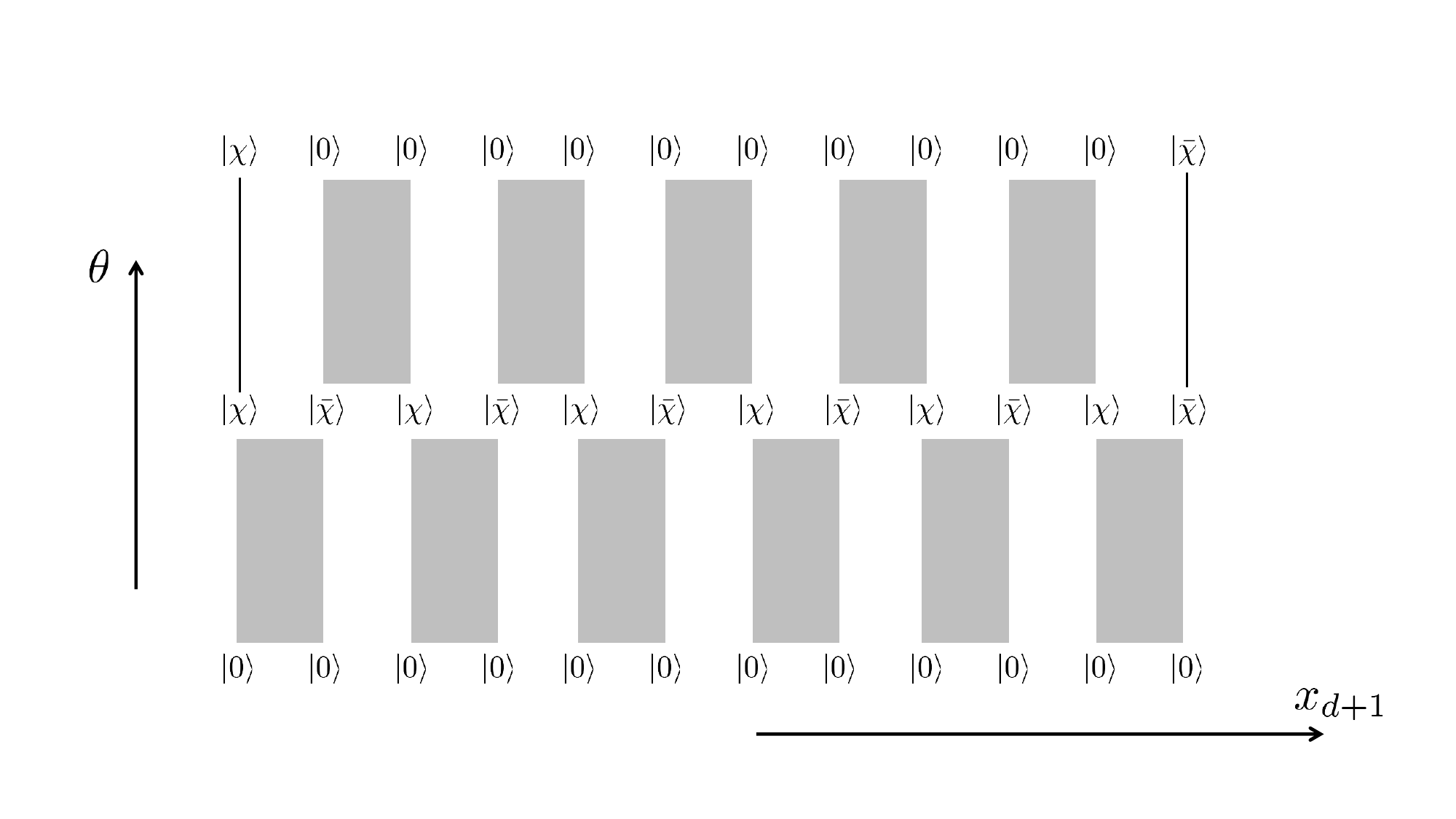}
\caption{Kitaev's canonical pump $E_d \to \Omega E_{d+1}$. }
\label{fig:Kitaev_pump}
\end{figure}

The $\Omega$-spectrum structure behind the invertible states is supported by the following canonical construction of the map $E_d \to \Omega E_{d+1}$, independent of the details of the system, from the following defining property of invertible states. 
For an invertible state $\ket{\chi}_d$ in $d$ dimensions, there is an invertible state $\ket{\bar \chi}_d$ such that the tensor product state $\ket{\chi}_d \otimes \ket{\bar \chi}_d$ is adiabatically equivalent to the tensor product state $\ket{0}_d \otimes \ket{0}_d$ of the trivial state $\ket{0}_d$. 
Let $\ket{0}_{d+1} = \bigotimes_{x \in \Z} \ket{x,0}_d$ be the trivial tensor product state in $(d+1)$ dimensions, where each state $\ket{x,0}_d$ is the copy of the trivial state of $d$ dimensions. 
In the first half period, the pair of trivial states at $2x-1$ and $2x$ are adiabatically deformed to the tensor product $\ket{2x-1,\chi}_d \otimes \ket{2x,\bar \chi}_d$, and in the second half, the pair of $2x$ and $2x+1$ sites are adiabatically deformed into the trivial states $\ket{2x,\chi}_d \otimes \ket{2x+1,\bar \chi}_d \sim \ket{2x,0}_d \otimes \ket{2x+1,0}_d$, resulting in an adiabatic cycle of $\Omega E_{d+1}$ labeled by $\ket{\chi}_d \in E_d$~\cite{Kit13}. 
We call this construction Kitaev's canonical pump. 
Clearly, for an open chain composed of even sites, the invertible state $\ket{\chi}_d$ and $\ket{\bar \chi}_d$ appear at each edge by a period of the adiabatic cycle (See Fig.~\ref{fig:Kitaev_pump}).  
Although a canonical construction of the inverse map $\Omega E_{d+1} \to E_d$ has not yet been known in lattice systems, $\Omega$-spectrum structure is consistent with various texture induced phenomena in invertible phases. 

It should be noted that the Thouless pump~\cite{Tho83}, where a $\Z$ charge pumped by an adiabatic cycle of 1D chain with $U(1)$ symmetry, is generalized to any discrete group symmetry and any spatial dimension. 
It is known that the topological invariant of the Thouless pump is the $U(1)$ phase winding of the charge polarization, the ground state expectation value of the twist operator~\cite{Res98}. 
On the one hand, for generic adiabatic cycles with discrete charge, such a physical geometric quantity of which the target space has a nontrivial first homotopy group labeled by the discrete charge is still unknown. 
To search such a geometric quantity is another motivation of this paper. 
We will see that in the cases where a non-chiral phase is pumped, the group cocycle $\omega_\theta \in Z^{d+1}(G,U(1))$ parameterized by the adiabatic parameter $\theta$ hosts the topological charge of cycles. 
Mathematically, this is understood from the isomorphism $H^d(G,U(1)) \cong H^{d+1}(G,\Z)$ from the Bockstein homomorphism associated with the short exact sequence $\R \to \Z \to U(1)$ of coefficients of group cohomologies. 
There, the group $(d+1)$-cocycle with $\Z$-coefficient is understood as the phase windings of the cocycle $\omega_\theta$. 
In Sec.~\ref{sec:any_dim}, we present an exactly solvable model of adiabatic cycles from the Bockstein homomorphism, in the flavor of Chen-Gu-Liu-Wen's construction~\cite{CGLW13}. 
Our construction turns out to be the same local unitary constructed in Ref.~\cite{RH17}.

For free fermions with or without translational invariance, the $\Omega$-spectrum structure is more tractable to formulate. 
For massive Dirac fermions ${\cal H} = \sum_{\mu=1}^d \g_\mu \p_\mu + M$ in $d$-dimension, the mass matrix $M$ is found to belong to the classifying space of the $K$-theory, this is nothing but the $\Omega$-spectrum of the $K$-theory~\cite{Kit09}. 
For translational invariant systems, the parameter-dependent adiabatic Hamiltonian ${\cal H}(\bm{k},s)$ is classified by the $K$-theory over the Bloch-momentum and parameter space. 
The topological classification of adiabatic cycles of the Hamiltonian ${\cal H}(\bk,s)$ is found to be the same as for Hamiltonians in the same symmetry class in one lower dimensions~\cite{TK10}. 

There are several related concepts and prior work for the adiabatic cycle in invertible states. 
In the Floquet SPT phases, periodically driven Hamiltonians are studied, and many protocols are known that pump an SPT phase at the boundary by a period. 
In the Floquet SPT phase, many-body localization is important to avoid thermalization. 
In this paper, we are interested in adiabatic cycles of the Hamiltonian, which closely overlaps with the topological classification of the Floquet SPT phase.
We should note that it was shown that a part of Floquet SPT phases are classified by the same classification of SPT phases in one lower dimensions (in addition to the static phases), which is the same conclusion from the $\Omega$-spectrum structure. 
There, the time-translation $\Z$ symmetry is introduced, which is generated by the Floquet unitary itself, and it is concluded that the Floquet SPTs are classified by the total symmetry group including $\Z$~\cite{EN16,KS16,PMV16,RH16,RH17}. 
In the context of field theory, 't Hooft anomalies are known to be classified by invertible phases in one higher dimensions. 
Adiabatic cycles of invertible theories correspond to one-parameter loops in non-anomalous or possibly anomalous theories on the boundary. 
For a nontrivial adiabatic cycle, an anomalous theory is pumped on the boundary, which leads to the existence of a phase transition at some adiabatic parameter. 
Related phenomena are discussed as the ``global inconsistency"~\cite{KT17}, the ``anomalies in the space of coupling constant"~\cite{CFLS20_1,CFLS20_2}, and the ``diabolical points in parameter space"~\cite{HKT20}. 

It is notable that adiabatic cycles with $U(1)$ symmetry in one-dimensional systems, i.e., Thouless pumps, have been realized in the cold atom system~\cite{nakajima2016topological,lohse2016thouless}.
As a physical system for realizing adiabatic cycles for generic finite group symmetries, the cold atom system should be a promising candidate.

Before moving on to the main part of the paper, we have some remarks. 
Firstly, the solvable models discussed in Secs.~\ref{sec:1DZ2}, \ref{sec:2DTRS} and \ref{sec:any_dim} are constructed by unitary transformations on reference Hamiltonians.
Therefore the spectrum does not change in the adiabatic time evolution, however, the ground state wave function does change, and in the presence of on-site symmetry of finite groups, the ``change'' of the wave function in one period is quantized in some sense, which is the phenomenon studied in this paper.
It is similar to the Berry phase, but the Berry phase is essentially a quantity for finite systems, i.e., in $0$-space dimension, but a new indicator is needed to characterize the change of the wave function in infinite systems of $d$-space dimensions, and we will discuss below that the group cocycle plays a role similar to the Berry phase. 
Secondly, although the models discussed in Secs.~\ref{sec:1DZ2}, \ref{sec:2DTRS} and \ref{sec:any_dim} are solvable and lacks generality, the physical phenomena demonstrated using the solvable model are characterized by topological invariants of the adiabatic cycle and are expected to be universal for adiabatic cycles in general.

The organization of this paper is as follows. 
In Sec.~\ref{sec:1DZ2}, we give a simple 1-dimensional model of the adiabatic pump with $\Z_2$ symmetry and study various tools to diagnose how the adiabatic cycle is nontrivial or not.
In Sec.~\ref{sec:MPS}, we study 1-dimensional adiabatic cycles from the matrix product state (MPS) description of 1-dimensional spin systems. 
In Sec~\ref{sec:2DTRS}, we give a simple 2-dimensional model of the adiabatic cycle with time-reversal symmetry (TRS), which is a model generalized from the Levin--Gu model~\cite{LG12}. 
In Sec.~\ref{sec:any_dim}, we present an exactly solvable model in $(d+1)$-dimensional adiabatic cycles from a given group cocycle in $d$-dimension. 
We again emphasize that the resulting model is the same local unitary constructed in Ref.~\cite{RH17}. 
We summarize this paper in Sec.~\ref{sec:sum}. 

Throughout this paper, we use $\theta$ as the adiabatic parameter with the period $2\pi$.
For a finite group $G$, we specify which $g \in G$ is unitary or antiunitary as a symmetry operation by a homomorphism $s: G \to \Z_2=\{1,-1\}$. 
``$d$-spatial dimension" and ``$d$-dimensional" are sometimes abbreviated as ``$d$D".

\section{Spin chain with $\Z_2$ symmetry
\label{sec:1DZ2}}
\subsection{A toy model}
\label{sec:1D_model}

\begin{figure}[!]
\centering
\includegraphics[width=\linewidth, trim=0cm 0cm 0cm 0cm]{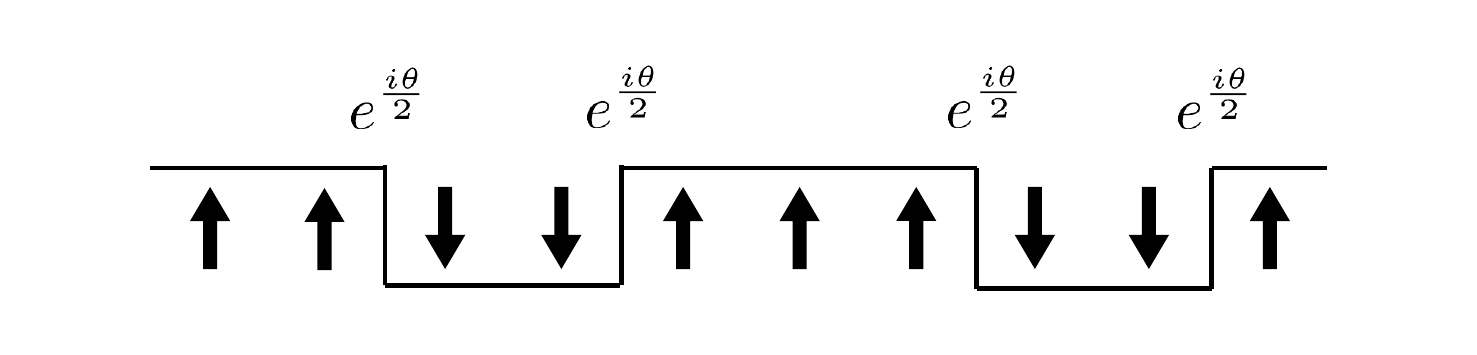}
\caption{The toy model (\ref{eq:1DZ2_GS}).}
\label{fig:model_1d}
\end{figure}

The trivial disordered phase with $\Z_2$ symmetry in spin 1/2 systems is described by the Hamiltonian 
\begin{align}
    H_0 = - \sum_{j \in \Z} \s^x_j. 
\end{align}
We define the $\Z_2$ symmetry operator by 
\begin{align}
 V = \prod_{j \in \Z} \s^x_j. 
 \label{eq:1D_Z2_Sym}
\end{align}
The ground state of $H_0$ is the fully polarized state $\ket{\Psi_0} = \ket{\cdots \to \to \cdots }$, and at the same time, it is written as the equal weight sum of the domain wall configurations 
\begin{align}
    \ket{\Psi_0} = \sum_{\{\s_j\}} \ket{\cdots \s_j \s_{j+1} \cdots}, 
\end{align}
up to a normalization factor, where $\s_j \in \{\ua,\da\}$. 
One can find this ground state can be modified by a $U(1)$ parameter $\theta$ while keeping $\Z_2$ symmetry as follows. 
Let $N_{\rm dw}$ be the number of domain walls, namely $N_{\rm dw}$ counts the states $\ua \da$ and $\da \ua$ in a configuration $\ket{\cdots \s_j \s_{j+1}\cdots }$. 
$N_{\rm dw}$ is defined explicitly by $N_{\rm dw} = \sum_{j \in \Z} (1-\s^z_j \s^z_{j+1})/2$ as an operator. 
We introduce the modified ground state by assigning the $U(1)$ phase $e^{i\theta/2}$ to each domain wall as in 
\begin{align}
    \Ket{\Psi_\theta} 
    =
    \sum_{\{\s_j\}} e^{\frac{i\theta}{2} N_{\rm dw}} \ket{\cdots \s_j \s_{j+1} \cdots}.
    \label{eq:1DZ2_GS}
\end{align}
This state is given by the local unitary transformation 
\begin{align}
    U_\theta = \prod_{j \in \Z} e^{\frac{i\theta}{2} \frac{1-\s^z_j \s^z_{j+1}}{2}} 
\label{eq:U_bulk}
\end{align}
on $\Ket{\Psi_0}$. 
Therefore, the Hamiltonian of which the ground state is $\Ket{\Psi_\theta}$ is given by 
\begin{align}
    H_\theta
    =
    U_\theta H_0 U_\theta^{-1}
    =
    - \sum_{j \in \Z} B^\theta_j, 
    \label{eq:H_bulk}
\end{align}
with 
\begin{align}
B_j^\theta 
&=
\s^x_j e^{\frac{i\theta}{2}\s^z_j (\s^z_{j-1}+\s^z_{j+1})} \nonumber \\
&= \frac{1+\cos\theta}{2} \s^x_j 
- \frac{1-\cos\theta}{2} \s^z_{j-1}\s^x_j \s^z_{j+1} \nonumber \\
&+ \frac{1}{2} \sin \theta (\s^z_{j-1} \s^y_j+\s^y_j \s^z_{j+1}). 
\end{align}
Notably, although the local terms $B^\theta_j$ of the adiabatic Hamiltonian $H_\theta$ is $2\pi$-periodic, the $2\pi$-periodicity of the ground state $\Ket{\Psi_\theta}$, or equivalently the local unitary $U_\theta$ holds only on the closed chain with the (anti)periodic boundary condition as the number $N_{\rm dw}$ of domain walls is even (odd). 

On an open chain with $L$ sites, one may define the local unitary 
\begin{align}
U_\theta = \prod_{j=1}^{L-1} e^{\frac{i\theta}{2} \frac{1-\s^z_j\s^z_{j+1}}{2}}. 
\label{eq:U1}
\end{align}
This local unitary is not $2\pi$-periodic at the boundary: 
There remain the $\Z_2$ charged operators as $\hat U_{2\pi} = \s^z_1 \s^z_L$. 
One can define another local unitary $\tilde U_\theta$ that is the same one as $U_\theta$ on closed chains. 
Let us consider 
\begin{align}
\tilde U_\theta = \prod_{j=1}^{L-1} e^{i\theta \frac{1+\s^z_j}{2}\frac{1-\s^z_{j+1}}{2}}. 
\label{eq:U2}
\end{align}
This is $2\pi$-periodic even for open chains, but $\Z_2$ symmetry is broken at the boundary.

\subsection{Open chain}
\label{sec:1D_open}
On an open chain with $L$ sites, we consider the Hamiltonian $H_\theta$ of the form 
\begin{align}
H_\theta
= H^{\rm bulk}_\theta + H^{\rm edge}_\theta, 
\end{align}
where the bulk part $H_\theta^{\rm bulk}$ is composed of local Hamiltonians $B^\theta_j$ and the sum runs over all sites in the interior of the chain. 
Namely, $H_\theta^{\rm bulk} = - \sum_{j=2}^{L-1} B^\theta_j$. 
The edge Hamiltonian $H_\theta^{\rm edge}$ is any local Hamiltonian that acts spins near the edge and is assumed to be small compared to the bulk gap. 
We first solve the bulk Hamiltonian $H_\theta^{\rm bulk}$ to get the degenerate ground states and discuss the effect of $H_\theta^{\rm edge}$ as the perturbation. 
$H_\theta^{\rm bulk}$ has four-fold ground state degeneracy because the edge spins $\s^z_1$ and $\s^z_L$ are not determined. 
The ground states are explicitly written as 
\begin{align}
\Ket{\Psi_\theta(\s_1,\s_L)} = \prod_{j=2}^{L-1}\frac{1+B^\theta_j}{2} \ket{\s_1 \ua\cdots \ua \s_L}, 
\label{eq:edge_psi}
\end{align}
where $\s_1,\s_L \in \{\ua,\da\}$. 
Here, $(1+B^\theta_j)/2$ are projection operators, and the reference states $\ket{\s_1 \ua\cdots \ua \s_L}$ are chosen not to vanish for the projections. 
It should be noted that the relative phases among ground states $\{\Ket{\Psi_\theta(\s_1,\s_L)}\}_{\s_1,\s_L \in \{\ua,\da\}}$ can not be fixed in general and can depend on $\theta$. 
We will discuss a phase choice depending on $\theta$ in Sec.~\ref{sec:comment_relative_phases}. 
The $\Z_2$ action on the degenerate ground states becomes $\theta$-dependent and explicitly written as 
\begin{align}
V \Ket{\Psi_\theta(\s_1,\s_L)} = e^{i\theta \frac{\s_1+\s_L}{2}} \Ket{\Psi_\theta(-\s_1,-\s_L)}, 
\end{align}
where $-\sigma_i$ denotes the opposite spin direction to $\sigma_i$. 
Introducing the Pauli matrices $\bar \s^\mu_1$ and $\bar \s^\mu_L$ for the degenerate ground states as in 
\begin{align}
&\bar \sigma_1^\mu
= \sum_{\sigma_L} \ket{\Psi_\theta(i,\sigma_L)} [\sigma^\mu]_{ij} \bra{\Psi_\theta(j,\sigma_L)}, \label{eq:bar_sigma_1}\\
&\bar \sigma_L^\mu
= \sum_{\sigma_1} \ket{\Psi_\theta(\sigma_1,i)} [\sigma^\mu]_{ij} \bra{\Psi_\theta(\sigma_1,j)},
\label{eq:bar_sigma_L}
\end{align}
we have the factorized form 
\begin{align}
P_\theta V P_\theta= v_1^\theta v^\theta_L
\label{eq:edge_Z2_1},  
\end{align}
with
\begin{align}
v_j^\theta = \bar \s^x_j e^{\frac{i\theta}{2} \bar \s^z_j} 
\label{eq:edge_Z2_2}
\end{align}
for $j=1$ and $L$. 
Here, 
\begin{align}
P_\theta = \sum_{\s_1,\s_L \in \{\ua,\da\}}
\ket{\Psi_\theta(\s_1,\s_L)}\bra{\Psi_\theta(\s_1,\s_L)}
\end{align}
is the projection onto the ground states. 
As will see later, one can define a $\Z_2$ invariant from the edge action $v^\theta_1$, which signals the nontrivial adiabatic cycle. 

Note that the gauge choice of $v_1^\theta$ and $v_L^\theta$ is not unique. 
To be precise, the $U(1)$ phase of $v_1^{\theta}$ is undetermined so that $v_1^\theta$ is a projective representation of $\Z_2$. 
The gauge choice shown in (\ref{eq:edge_Z2_2}) is chosen such that $(v_j^\theta)^2 = 1$ holds. 
However, the gauge choice $(\ref{eq:edge_Z2_2})$ breaks the $2\pi$-periodicity. 
Another gauge choice is 
\begin{align}
    \tilde v_1^\theta = \bar \s^x_1 e^{i\theta \frac{1+\bar \s^z_1}{2}}, 
    \qquad
    \tilde v_L^\theta = \bar \s^x_L e^{-i\theta \frac{1-\bar \s^z_L}{2}}.
    \label{eq:edge_Z2_3}
\end{align}
This maintains the $2\pi$-periodicity, but breaks the $\Z_2$-ness as it obeys $(\tilde v^\theta_1)^2 = e^{i\theta}$. 
We note that $\tilde v^\theta_1$ and $\tilde v^\theta_L$ are still projective representations of $\Z_2$. 

Let us consider some edge Hamiltonians below.

\subsubsection{Edge Hamiltonian with $\Z_2$ symmetry and without $2\pi$-periodicity}
We first consider the edge Hamiltonian with $\Z_2$ symmetry but without the $2\pi$-periodicity. 
Such an edge Hamiltonian is given by, for example, 
\begin{align}
H_\theta^{\rm edge} 
&= - \lambda U_\theta (\s^x_1+\s^x_L) U_\theta^{-1} \nonumber \\
&= - \lambda \s^x_1 e^{\frac{i\theta}{2} \s^z_1\s^z_2} -  \lambda \s^x_L e^{\frac{i\theta}{2} \s^z_{L-1}\s^z_L} 
\end{align}
with $U_\theta$ the local unitary introduced in (\ref{eq:U1}). 
$H^\theta_{\rm edge}$ is not $2\pi$-periodic as $U_\theta$ so. 
The total Hamiltonian is still composed of commuting local terms, implying that the eigenstates of the edge effective Hamiltonian $P_\theta H_\theta^{\rm edge} P_\theta$ are exact ones. 
The edge effective Hamiltonian reads as 
\begin{align}
P_\theta H_\theta^{\rm edge} P_\theta
= - \bar \s^x_1 e^{\frac{i\theta}{2} \bar \s^z_1} - \bar \s^x_L e^{\frac{i\theta}{2} \bar \s^z_L}, 
\end{align}
and the ground state is given by 
\begin{align}
\Ket{\Psi_\theta} \sim 
\begin{pmatrix}
1\\
e^{\frac{i\theta}{2}}
\end{pmatrix}_{\bar \s_1} 
\otimes 
\begin{pmatrix}
1\\
e^{\frac{i\theta}{2}}
\end{pmatrix}_{\bar \s_L}. 
\end{align}
Note that $\Ket{\Psi_\theta}$ is not $2\pi$-periodic as $H_\theta^{\rm edge}$ explicitly breaks it, and the $\Z_2$ charge at the edge can be constant as $v_1^\theta (1,e^{\frac{i\theta}{2}})_{\bar \s_1}^T = 1$. 

\subsubsection{Edge Hamiltonian without $\Z_2$ symmetry and with $2\pi$-periodicity }
Now consider the opposite case where $\Z_2$ is explicitly broken but the $2\pi$-periodicity is possessed. 
An example of such an edge Hamiltonian is given by 
\begin{align}
H_\theta^{\rm edge} 
&= -\lambda \tilde U_\theta (\s^x_1+\s^x_L) [\tilde U_\theta]^{-1} \nonumber \\
&= - \lambda \s^x_1 e^{-i\theta \s^z_1 \frac{1-\s^z_2}{2}} - \lambda \s^x_L e^{i\theta \s^z_L \frac{1+\s^z_{L-1}}{2}}, 
\end{align}
with $U_\theta$ the local unitary introduced in (\ref{eq:U2}). 
The edge effective Hamiltonian is 
\begin{align}
P_\theta H_\theta^{\rm edge} P_\theta = - \lambda\bar \s^x_1 - \lambda \bar \s^x_L e^{i\theta \bar \s^z_L}, 
\end{align}
and the ground state is 
\begin{align}
\Ket{\Psi} \sim 
\begin{pmatrix}
1\\
1\\
\end{pmatrix}_{\bar \s_1} 
\otimes 
\begin{pmatrix}
1\\
e^{i\theta}
\end{pmatrix}_{\bar \s_L}. 
\end{align}
This is $2\pi$-periodic, but does not have $\Z_2$ symmetry. 

\subsubsection{Edge Hamiltonian with $\Z_2$ symmetry and $2\pi$-periodicity}
\label{sec:1DZ2_edge} 
An example of edge Hamiltonian satisfying both $\Z_2$ symmetry and the $2\pi$-periodicity is a constant one 
\begin{align}
H_\theta^{\rm edge}= -\lambda (\s^x_1+\s^x_L). 
\label{eq:H_edge_2}
\end{align}
$H_\theta^{\rm edge}$ is not closed on the ground state manifold as $H_\theta^{\rm edge}$ does not commute with the bulk one $H_\theta^{\rm bulk}$. 
The first-order effective edge Hamiltonian is given by 
\begin{align}
&P_\theta H_\theta^{\rm edge} P_\theta  \nonumber \\
&=
- \lambda \cos \frac{\theta}{2} 
\left( 
e^{-\frac{i\theta}{4}\bar \s^z_1}\bar\s^x_1e^{\frac{i\theta}{4}\bar \s^z_1}
+e^{-\frac{i\theta}{4}\bar \s^z_L}\bar\s^x_Le^{\frac{i\theta}{4}\bar \s^z_L}\right).
\end{align}
There is a level crossing, and the ground state is degenerate at $\theta=\pi$. 
In other words, the ground state can not be unique for all $\theta \in [0,2\pi]$. 
The lowest two eigenstates of $P_\theta H_\theta^{\rm edge} P_\theta$ are given by 
\begin{align}
\Ket{\Psi^\pm_\theta}
\sim \begin{pmatrix}
1\\
\pm e^{\frac{i\theta}{2}}\\
\end{pmatrix}_{\bar \s_1} \otimes 
\begin{pmatrix}
1\\
\pm e^{\frac{i\theta}{2}}\\
\end{pmatrix}_{\bar \s_L}. 
\end{align}
Two states $\Ket{\Psi^+_\theta}$ and $\Ket{\Psi^-_\theta}$ are interchanged by a period as $\Ket{\Psi^+_{\theta+2\pi}} = \Ket{\Psi^-_\theta}$. 
Although the effective edge Hamiltonian $P_\theta H_\theta^{\rm edge} P_\theta$ is $2\pi$-periodic, the lowest two states can be regarded as a single state with the $4\pi$-periodicity. 

\begin{figure}[!]
\centering
\includegraphics[width=\linewidth, trim=0cm 0cm 0cm 0cm]{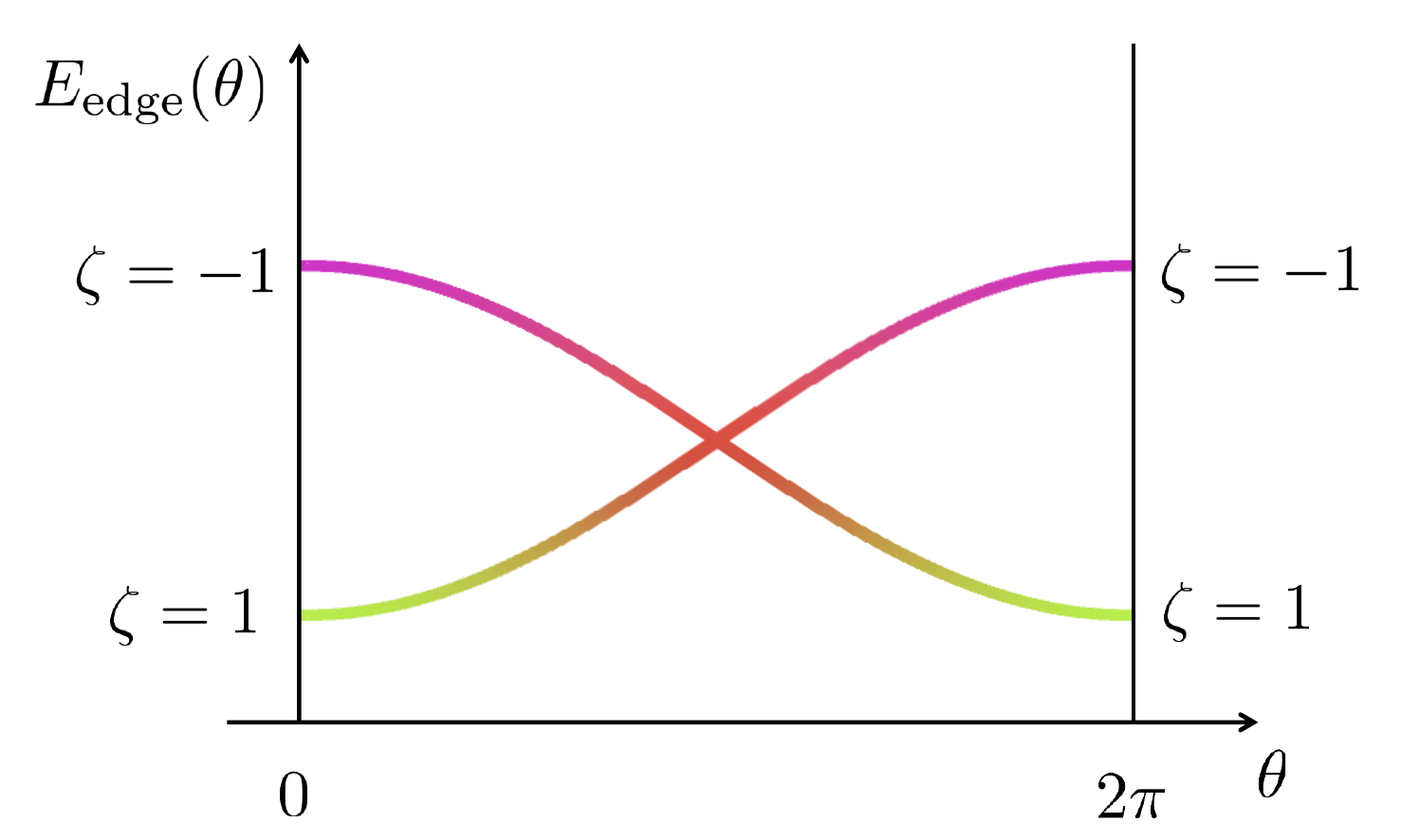}
\caption{Edge spectrum of the edge Hamiltonian with $\Z_2$ symmetry and the $2\pi$-periodicity. 
$\zeta$s represent the eigenvalues of the edge $\Z_2$ symmetry (\ref{eq:edge_Z2_3}), a one-parameter family of projective representations of $\Z_2$.}
\label{fig:edge_spectrum}
\end{figure}

Let us focus on the states 
\begin{align*}
    \ket{\psi^\pm_\theta} \sim 
    \begin{pmatrix}
    1\\
    \pm e^{\frac{i\theta}{2}}\\
    \end{pmatrix}_{\bar \sigma_1} 
\end{align*}
at the left edge. 
To have a continuous eigenvalue of $\Z_2$ action, we employ the gauge choice (\ref{eq:edge_Z2_3}). 
We find that the eigenvalue of $\tilde v^\theta_1$ is also $4\pi$-periodic as $\tilde v^\theta_1 \ket{\psi^\pm_\theta} = \pm e^{i\theta/2}$. 
See Fig.~\ref{fig:edge_spectrum}. 
This nature of $4\pi$-periodicity is the origin of the unavoidable level crossing, as discussed below. 


\subsection{Projective representation and $\Z_2$ invariant}
\label{sec:1D_projective_rep_Z2_inv}
In Sec.~\ref{sec:1DZ2_edge}, we saw that there is a level crossing in the edge spectrum for a $\Z_2$ symmetric and $2\pi$-periodic edge Hamiltonian in the first-order calculation. 
One can show that the level crossing is a consequence of the nontrivial cycle of the edge $\Z_2$ action (\ref{eq:edge_Z2_2}). 

Since the $U(1)$ phase of the $\Z_2$ action $v_1^\theta$ at the left edge is unfixed in the expression (\ref{eq:edge_Z2_1}), the matrix $v_1^\theta$ should be considered as a projective representation of $\Z_2$. 
For 1D SPT phases in spin systems, the nontrivial factor system of the projective representation of symmetry group $G$ signals nontrivial SPT phases~\cite{PETO12, CGW11, SPC11}. 
On the one hand, since the $\Z_2$ group has no nontrivial factor system as $H^2(\Z_2,U(1))=0$, the projective representation $v^\theta_1$ belongs to the trivial projective representation. 

Nevertheless, as a cycle of projective representation, the factor system of $v^\theta_1$ is nontrivial. 
To see this, we first note that a generic $2\pi$-periodic projective representation of $\Z_2$, $u_\theta$ with  $(u_\theta)^2 \sim {\bf 1}$, defines a $\Z_2$ invariant.
Let $\omega_\theta \in U(1)$ be the two-cocycle (factor system) defined as $(u_\theta)^2 = \omega_\theta {\bf 1}$. 
The $\Z_2$ invariant is defined by 
\begin{align}
\nu = \frac{1}{2\pi i} \oint d \log \omega_\theta \quad {\rm mod\ }2. 
\label{eq:Z2inv}
\end{align}
Even integers in $\nu$s are meaningless since if one replace the $U(1)$ phase of $u_\theta$ by $u_\theta \mapsto e^{in \theta} u_\theta$ with an integer $n$, the invariant $\nu$ changes by $2n$. 
The edge $\Z_2$ action (\ref{eq:edge_Z2_3}) has the nontrivial $\Z_2$ invariant $\nu \equiv 1$. 

The level crossing is the consequence of $\nu \equiv 1$: 
Suppose that the ground state is unique for all $\theta$, and $\Z_2$ symmetry is unbroken. 
Then the edge projective representation of $\Z_2$ is a one-dimensional representation $u_\theta = e^{i \alpha(\theta)}$ in which the $\Z_2$ invariant $\nu$ is trivial due to the $2\pi$-periodicity of $e^{i\alpha(\theta)}$. 
Therefore, we conclude that the nontrivial $\Z_2$ invariant $\nu \equiv 1$ implies the ground state degeneracy at some $\theta$.

\subsection{Comment on the relative phases of ground states}
\label{sec:comment_relative_phases}
As commented in Sec.~\ref{sec:1D_open}, the relative phase among the degenerate ground states $\Ket{\Psi_\theta(\s_1,\s_L)}$ can be chosen such that they explicitly depend on $\theta$. 
For example, let $\Ket{\Psi_\theta(\s_1,\s_L)}'$ be the basis obtained by the non-local unitary transformation on $\Ket{\Psi_\theta(\s_1,\s_L)}$ as in 
\begin{align}
\Ket{\Psi_\theta(\s_1,\s_L)}' 
= e^{-i\theta \frac{1-\s^z_1}{2}\frac{1-\s^z_L}{2}} \Ket{\Psi_\theta(\s_1,\s_L)}. 
\end{align}
We introduce the effective edge spin operators $\bar \sigma'^\mu_1$ and $\bar\sigma'^\mu_L$ in the same way as (\ref{eq:bar_sigma_1}) and (\ref{eq:bar_sigma_L}) on the basis $\Ket{\Psi_\theta(\s_1,\s_L)}'$. 
The effective $\Z_2$ action reads a constant $P_\theta VP_\theta = \bar \s'^x_1 \bar \s'^x_L$.
Correspondingly, the effective edge Hamiltonian becomes non-local. 
For example, the edge Hamiltonian (\ref{eq:H_edge_2}) becomes
\begin{align}
P_\theta H_\theta^{\rm edge} P_\theta \nonumber
&=
-\lambda \cos \frac{\theta}{2} \Big(
e^{\frac{i\theta}{4}\bar \s'^z_1 \bar \s'^z_L}\bar\s'^x_1e^{-\frac{i\theta}{4}\bar \s'^z_1 \bar \s'^z_L} \\
&\quad +e^{\frac{i\theta}{4}\bar \s'^z_1 \bar \s'^z_L}\bar\s'^x_1e^{-\frac{i\theta}{4}\bar \s'^z_L \bar \s'^z_L}
\Big). 
\end{align}
On the basis $\Ket{\Psi_\theta(\s_1,\s_L)}'$, one can not extract the nontrivial $\Z_2$ cycle from the effective edge symmetry action. 
Therefore, the locality of the phase choice of the ground states $\Ket{\Psi_\theta(\s_1,\s_L)}$ is crucial to define the $\Z_2$ invariant (\ref{eq:Z2inv}).

\subsection{$\Z_2$ charge trapped on a spatial texture}
\label{sec:1D_texture}
Another way to detect the nontriviality of the adiabatic cycle $H_\theta$ is to measure the symmetry charge of the ground state under a spatial texture in which $\theta$ slowly varies in the space from $0$ to $2\pi$.  

\begin{figure}[!]
\centering
\includegraphics[width=\linewidth, trim=0cm 0cm 0cm 0cm]{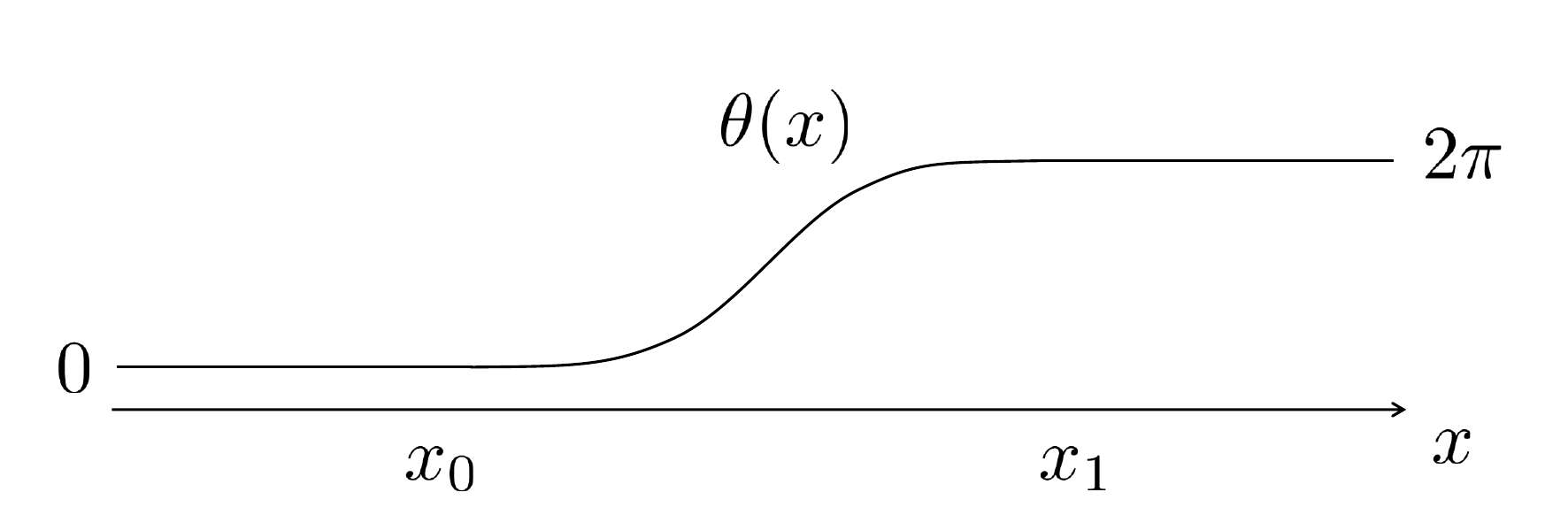}
\caption{A function $\theta$ varying in space form $0$ to $2\pi$.}
\label{fig:texture_func}
\end{figure}

Let $\theta(x)$ be a $\R$-valued smooth function such that 
\begin{align}
\theta(x) = \left\{ \begin{array}{ll}
0 & (x \leq x_0), \\
2\pi & (x \geq x_1). 
\end{array}
\right. 
\label{eq:theta_func}
\end{align}
Here, $x_0$ and $x_1$ are positions with $x_0<x_1$, and $|x_1-x_0|$ is large enough to the inverse of the energy gap. 
See Fig.~\ref{fig:texture_func} for a function $\theta(x)$. 
The Hamiltonian with a spatial texture is of a form 
\begin{align}
H_{\rm texture} = - \sum_{j} B_j^{\theta(j)}. 
 \label{eq:H_tx}
\end{align}
We claim that the ratio of the $\Z_2$ charges of the ground states between $H_0$ and $H_{\rm texture}$ is the $\Z_2$ invariant to detect if a given cycle $H_\theta$ is nontrivial or not. 
Although this strategy can be applied to any adiabatic Hamiltonian $H_\theta$ with translational invariance, we show that for models that obtained by the local unitary transformation, one has a Hamiltonian which approximates the texture Hamiltonian $H_{\rm texture}$ of the form (\ref{eq:H_tx}), as explained below. 

By using the local unitary (\ref{eq:U2}), the texture Hamiltonian is given by $\tilde H_{\rm texture} = \tilde U_{\rm twist} H_0 [U_{\rm twist}]^{-1}$ with 
\begin{align}
\tilde U_{\rm twist} = \prod_j e^{i \theta(j) \frac{1+\s^z_j}{2}\frac{1-\s^z_{j+1}}{2}}.
\end{align}
However, since $\tilde U_{\rm twist}$ breaks the $\Z_2$ symmetry slightly as $V \tilde U_{\rm twist} V^{-1} = \tilde U_{\rm twist} e^{-i \sum_j \s^z_j \frac{\theta_j-\theta_{j-1}}{2}}$, the ground state expectation value of the $\Z_2$ operator $V$ is quantized only in the thermodynamic limit. 
We do not describe this type of the twist operator in the details. 

Instead, we consider the twist operator 
\begin{align}
U_{\rm twist} = \prod_j e^{\frac{i\theta(j)}{2} \frac{1-\s^z_j\s^z_{j+1}}{2}}
\end{align}
which preserves $\Z_2$ symmetry $V U_{\rm twist} V^{-1} = U_{\rm twist}$. 
We note that $U_{\rm twist}$ has the support only on $x_0-1 \leq j \leq x_1+1$ even if the unitary transformation $e^{\frac{i\theta}{2} \frac{1-\s^z_j\s^z_{j+1}}{2}}$ per site is not $2\pi$-periodic as it shows $e^{i\pi \frac{1-\s^z_j\s^z_{j+1}}{2}} = \s^z_j \s^z_{j+1}$. 
This is because the contributions from nearest neighbor sites are canceled out for $j>x_1+1$, resulting in that the local terms of the texture Hamiltonian $U_{\rm twist} H_0 U_{\rm twist}^{-1}$ is unchanged for $j <x_0-1$ and $j>x_1+1$. 

However, remarkably, the twist operator $U_{\rm twist}$ does not work to give a smooth texture Hamiltonian for closed chains with the periodic boundary condition where the $1$ and $L+1$ sites are identified. 
To see this, let us try to apply the following trial twist operator on the closed chain 
\begin{align}
U_{\rm twist, trial}^{S^1} 
= \prod_{j=1}^L e^{\frac{i\theta(j)}{2}\frac{1-\s^z_j \s^z_{j+1}}{2}} 
\end{align}
to get the texture Hamiltonian $H_{\rm texture} = U^{S^1}_{\rm twist, trial} H_0 [U_{\rm twist,trial}^{S_1}]^{-1} = - \sum_{j=1}^N B_j^{\rm tx}$. 
The local terms read as 
\begin{align}
B^{\rm tx}_j 
&= U_{\rm twist, trial}^{S^1} \s^x_j [U_{\rm twist, trial}^{S^1}]^{-1} \nonumber \\
&=
\cos \frac{\theta(j-1)}{2} \cos \frac{\theta(j)}{2} \s^x_j\nonumber \\
&- \sin \frac{\theta(j-1)}{2} \sin \frac{\theta(j)}{2} \s^z_{j-1}\s^x_j \s^z_{j+1}\nonumber\\
&+ \sin \frac{\theta(j-1)}{2} \cos \frac{\theta(j)}{2} \s^z_{j-1} \s^y_j \nonumber\\
&+ \cos \frac{\theta(j-1)}{2} \sin \frac{\theta(j)}{2} \s^y_j \s^z_{j+1}
\end{align}
for $j=2,\dots, L$, and 
\begin{align}
B^{\rm tx}_1 
&= U_{\rm twist, trial}^{S^1} \s^x_1 [U_{\rm twist, trial}^{S^1}]^{-1}\nonumber\\
&=
\cos \frac{\theta(L)}{2} \cos \frac{\theta(1)}{2} \s^x_1\nonumber\\
&- \sin \frac{\theta(L)}{2} \sin \frac{\theta(1)}{2} \s^z_{L}\s^x_1 \s^z_{2} \nonumber\\
&+ \sin \frac{\theta(L)}{2} \cos \frac{\theta(1)}{2} \s^z_{L} \s^y_1 \nonumber\\
&+ \cos \frac{\theta(L)}{2} \sin \frac{\theta(1)}{2} \s^y_1 \s^z_{2}.
\end{align}
Since $\theta(N)=2\pi$, $B^{\rm tx}_j$s are singular at site $1$ and are not smooth. 
To compensate for this discrepancy, the twist operator needs to be modified for closed chains by inserting the $\Z_2$ charged operator at $j=1$ as in 
\begin{align}
U^{S^1}_{\rm twist}
:= 
\s^z_1 \prod_{j=1}^L e^{\frac{i\theta(j)}{2}\frac{1-\s^z_j \s^z_{j+1}}{2}}.
\label{eq:twist_op_s1}
\end{align}
With this twist operator, we have the texture Hamiltonian $H_{\rm texture} = U^{S^1}_{\rm twist} H_0 [U_{\rm twist}^{S_1}]^{-1}$ which smoothly varies in the closed chain and have a unit winding of $\theta$. 

Now let us evaluate the ground state expectation value of the texture Hamiltonian $H_{\rm texture}$. 
No explicit calculation is needed. 
The ground state $\Ket{\Psi_{\rm texture}}$ of $H_{\rm texture}$ is given by the unitary transformation $\Ket{\Psi_{\rm texture}} = U^{S^1}_{\rm twist} \Ket{\Psi_0}$. 
From the algebraic relation 
\begin{align}
V U^{S^1}_{\rm twist} V^{-1} = - U^{S^1}_{\rm twist}, 
\end{align}
where the factor (-1) comes from the charged operator $\s^z_1$, we conclude that a spatial texture of the adiabatic Hamiltonian (\ref{eq:H_bulk}) has the nontrivial $\Z_2$ charge. 

In Sec.~\ref{sec:texture_SPT}, we generalize the prescription here to adiabatic cycles in any spatial dimensions for an exactly solvable model.

\subsection{Berry phase}
In the previous section, we considered the spatial texture of the adiabatic Hamiltonian. 
In this section, we consider an alternative one, the temporal texture with the twisted boundary condition. 
Let $\Ket{\Psi_\theta^\s} (\Ket{\Psi_\theta^0})$ be the family of the ground states of the adiabatic Hamiltonian $H_\theta^\s (H_\theta^0)$ for the twisted boundary condition by $\Z_2$ symmetry (for the periodic boundary condition, resp.). 
For the spin system introduced in Sec.~\ref{sec:1D_model}, the twisted boundary condition is defined by the identification rule 
\begin{align}
\bm{\s}_{j+L} = V \bm{\s}_j V^{-1}
\end{align}
for the spin operators. 
Let $e^{i\g_0}$ and $e^{i\g_\s}$ be the Berry phases for the periodic and boundary conditions, respectively. 
We claim that the ratio of the Berry phases 
\begin{align}
e^{i\gamma_\s}/e^{i\gamma_0}
\label{eq:ratio_Berry_phase}
\end{align}
is quantized to a $\Z_2$ value in the thermodynamic limit and serves as the $\Z_2$ invariant of the adiabatic cycle. 

For the toy model (\ref{eq:H_bulk}), the Berry phase is computed as follows. 
The ground state for the periodic/twisted boundary condition is given by 
\begin{align}
    \ket{\Psi^{0/\s}_\theta}
    =U_\theta^{0/\s} \ket{\Psi_0},
\end{align}
where $\ket{\Psi_0}$ is the fully polarized state $\ket{\Psi_0}=\ket{\to\to\cdots}$, and $U^{0/\s}_\theta = e^{\frac{i\theta}{2} N_{\rm dw}^{0/\s}}$ is the local unitary 
with 
\begin{align}
N_{\rm dw}^{0/\s}=\sum_{j=1}^{L-1} \frac{1-\s^z_j\s^z_{j+1}}{2}+\frac{1\mp\s^z_L\s^z_1}{2}
\end{align}
the operator counting domain walls for the periodic/twisted boundary condition. 
Since $U^{0/\s}_{2\pi}=\pm {\rm Id}$ holds as an operator, the ground state satisfies the boundary condition $\ket{\Psi^{0/\s}_{2\pi}} = \pm \ket{\Psi^{0/\s}_0}$, and it contributes the Berry phase by $e^{i \pi}$ for the twisted boundary condition. 
For the both boundary conditions, the contribution from the integral of the Berry connection to the Berry phase results in a common value 
\begin{align}
e^{\int_0^{2\pi} \braket{\Psi^{0/\s}_\theta|d_\theta \Psi^{0/\s}_\theta}}
=
e^{i\pi \braket{\Psi_0|N^\s_{\rm dw}|\Psi_0}}
=i^L.
\end{align}
In sum, the Berry phases are $e^{i\gamma_\s}=-1$ and $e^{i\gamma_0}=1$, and therefore, 
the adiabatic Hamiltonian (\ref{eq:H_bulk}) shows a nontrivial ratio $(-1)$ of the Berry phases. 

Note that the ratio (\ref{eq:ratio_Berry_phase}) is generally not quantized when there is no symmetry to quantize the Berry phase.
There is an example of a model in which the ratio (\ref{eq:ratio_Berry_phase}) is quantized only in the thermodynamic limit~\cite{Ohyama}.

\subsection{Duality transformations}
In the presence of $\Z_2$ onsite symmetry, one can apply the Kramers--Wanner and the Jordan--Wigner duality maps to get dual Hamiltonians. 
It should be instructive to see dual models of (\ref{eq:H_bulk}). 

\subsubsection{Kramers--Wannier map}
We apply the dictionary of the Kramers--Wannier duality map 
\begin{align}
&\sigma^x_j \mapsto \tau^y_j \tau^y_{j+1}, \label{eq:KW_1}\\
&\sigma^z_j \sigma^z_{j+1} \mapsto \tau^z_{j+1}
\label{eq:KW_2}
\end{align}
to the model (\ref{eq:H_bulk}). 
We have the dual trivial Hamiltonian $H^{\rm KW}_0$ and the local unitary $U^{\rm KW}_\theta$ as follows. 
\begin{align}
&H^{\rm KW}_0=-\sum_j \tau^y_j \tau^y_{j+1}, \\
&U^{\rm KW}_\theta=\prod_j e^{\frac{i\theta}{2} \frac{1-\tau^z_j}{2}}.
\end{align}
We also have the $\Z_2$ onsite symmetry (Wilson line) $W$ in the dual model 
\begin{align}
W = \prod_j \tau^x_j. 
\end{align}
$H^{\rm KW}_0$ is the Ising Hamiltonian and the dual local unitary $U^{\rm KW}_{\theta}$ can be seen as assigning the $U(1)$ phase $e^{\frac{i\theta}{2}}$ to the charged objects $\tau^z_j$ for the $\Z_2$ symmetry $W$. 
To be concrete, the dual adiabatic Hamiltonian is still the Ising model but the spin axis is rotated by $\theta/2$ around the $z$-axis as in 
\begin{align}
H^{\rm KW}_\theta
&= U_\theta^{\rm KW} H^{\rm KW}_0[U_\theta^{\rm KW}]^{-1} \nonumber\\
&= - \sum_j \tau^y_j(\frac{\theta}{2}) \tau^y_j(\frac{\theta}{2}), 
\label{eq:1D_Hamiltonian_KW}
\end{align}
where $\bm{\tau}_j(\phi) = e^{-i\phi \frac{\tau^z_j}{2}} \bm{\tau}_j e^{i\phi \frac{\tau^z_j}{2}}$. 

Since the adiabatic Hamiltonian (\ref{eq:1D_Hamiltonian_KW}) is an Ising model for all $\theta$, the ground state is in a spontaneous symmetry broken phase. 
For the closed chain with the periodic boundary condition, the ground state has two-fold degeneracy and is spanned by the cat states $\ket{\pm(\theta)}$ characterized by $\tau^y_j(\theta/2) \equiv \pm 1$ for all $j$. 
Although the spin operator $\bm{\tau}_j(\theta/2)$ is not $2\pi$-periodic but $4\pi$-periodic, the Ising term $\tau^y_j(\frac{\theta}{2}) \tau^y_j(\frac{\theta}{2})$ is $2\pi$-periodic, implying that during a period the two cat states are exchanged. 

In this paper, we do not study the adiabatic cycles in spontaneous symmetry broken phases anymore.
We should note that the Floquet drives in spontaneous symmetry broken phases were studied in Ref.~\cite{KS16_2}.

\subsubsection{Jordan--Wigner map}
Let $a_j,a_j^\dag$ be complex fermion creation and annihilation operators at site $j$. 
By introducing the Majorana fermion operators $c_{2j-1}, c_{2j}$ by 
\begin{align}
&c_{2j-1}=-i(a_j-a^\dag_j), \\
&c_{2j} = a_j+a_j^\dag, 
\end{align}
the Jordan--Wigner transformation for the $\Z_2$ symmetry $V$ is given by 
\begin{align}
&\s^y_j = c_{2j} \prod_{i<j} (ic_{2i-1}c_{2i}) , \label{eq:JW_1}\\
&\s^z_j = c_{2j-1} \prod_{i<j} (ic_{2i-1}c_{2i}), \label{eq:JW_2}\\
&\s^x_j = i c_{2j-1}c_{2j}. \label{eq:JW_3}
\end{align}
Applying the Jordan--Wigner map to the model (\ref{eq:H_bulk}), we have the dual Hamiltonian $H^{\rm JW}_0$ and the local unitary $U^{\rm JW}_\theta$ as 
\begin{align}
H^{\rm JW}_0 
&= - \sum_j (ic_{2j-1}c_{2j}) \nonumber \\
&= -\sum_j (1-2a^\dag_ja_j), 
\end{align}
\begin{align}
U^{\rm JW}_\theta =\prod_j e^{\frac{i\theta}{2} \frac{1-ic_{2j}c_{2j+1}}{2}}.
\end{align}
Importantly, the local unitary $U^{\rm JW}_\theta$ does not give a $U(1)$ phase on the local $U(1)$ charge of the complex fermions $a_j^\dag$, but on the complex fermions living in bonds. 
The dual adiabatic Hamiltonian is 
\begin{align}
H^{\rm JW}_\theta=-\sum_j B^{{\rm JW},\theta}_j 
\end{align}
with 
\begin{align}
B^{{\rm JW},\theta}_j
&=\frac{1+\cos \theta}{2} (1-2a^\dag_j a_j) \nonumber\\
&- \frac{1-\cos \theta}{2} (a_{j-1}+a_{j-1}^\dag)(a_{j+1}-a^\dag_{j+1}) \nonumber\\
&+ i \sin \theta (a_ja_{j+1}+a_j^\dag a_{j+1}^\dag). 
\end{align}
The adiabatic cycle $H^{\rm JW}_\theta$ is supposed to show a nontrivial fermion parity pump. 
Since the state at $\theta=0$ is the vacuum, the fermion parity pump of $H^{\rm JW}_\theta$ is realized in a $\Z_2$-trivial superconductor which has no edge Majorana modes. 

\subsubsection{Kramer--Wannier and Jordan--Wigner map}
The final duality map is the successive map of the Kramers--Wannier map followed by the Jordan--Wigner transformation. 
This is same as the half lattice transformation $c_j \mapsto c_{j+1}$ of Majorana fermions. 
The resulting model at $\theta=0$ is the zero-correlation limit of the Kitaev chain~\cite{Kit01}
\begin{align}
H^{\rm KWJW}_0 
= - \sum_j (ic_{2j}c_{2j+1}). 
\end{align}
The mapped local unitary is the $e^{i\theta/2}$ phase rotation of the complex fermions 
\begin{align}
U^{\rm KWJW}_\theta
= 
\prod_j e^{\frac{i\theta}{2} a^\dag_j a_j}. 
\end{align}
Thus, the dual adiabatic model $H^{\rm KWJW}_\theta$ is the $2\pi$-phase rotation of the superconducting gap function 
\begin{align}
H^{\rm KWJW}_\theta
&=
\sum_j(
-a^\dag_ja_{j+1}-a^\dag_{j+1}a_j \nonumber \\
&+e^{i\theta}a^\dag_j a^\dag_{j+1}+e^{-i\theta}a_{j+1}a_j).
\end{align}
It is well known that the $2\pi$-phase rotation of a $\Z_2$-nontrivial superconductor gives rise to the fermion parity pump~\cite{Kit01}.

\subsection{Other models}
We present other models of the $\Z_2$ charge pump in spin chains. 

\subsubsection{The cluster Hamiltonian}
Let us consider the cluster Hamiltonian~\cite{BR01}
\begin{align}
H_0 = - \sum_j \sigma^z_{j-1}\sigma^x_j \sigma^z_{j+1}. 
\end{align}
This model has $\Z_2$ symmetry, of which the symmetry operator is the same form as (\ref{eq:1D_Z2_Sym}). 
The cluster Hamiltonian can be modified while keeping the $\Z_2$ symmetry by the local unitary 
\begin{align}
U_\theta = \prod_j e^{\frac{i\theta}{2} \frac{1-\sigma^x_j}{2}}.
\label{eq:1D_U_2}
\end{align}
We consider the adiabatic Hamiltonian $H_\theta= U_\theta H_0 U_\theta^{-1}=-\sum_j B^\theta_j$ with $B^\theta_j =\sigma^z_{j-1}(\frac{\theta}{2}) \sigma^x_j \sigma^z_{j+1}(\frac{\theta}{2})$ and $\bm{\sigma}_j(\phi)=e^{-i\phi \frac{\sigma^x_j}{2}} \bm{\sigma}_j e^{i\phi\frac{\sigma^x_j}{2}}$. 
Unlike the local unitary (\ref{eq:U_bulk}) discussed in Sec.~\ref{sec:1D_model}, the local unitary (\ref{eq:1D_U_2}) for a period, $U_{2\pi}$, is not the identity, but coincides with the $\Z_2$ symmetry operator $U_{2\pi}=V$. 
Since the Hamiltonian $H_0$ is $\Z_2$ symmetric, $H_\theta$ becomes $2\pi$-periodic.

Let us consider the model $H_\theta$ on an open chain. 
The total Hamiltonian is of a form $H_{\rm bulk}^\theta + H^\theta_{\rm edge}$ where $H^\theta_{\rm bulk} = - \sum_{j=2}^{N-1} B^\theta_j$. 
Since $B^\theta_j$s are commuted each other, the ground state of $H^\theta_{\rm bulk}$ is given by imposing $B^\theta_j=1$ for $j=2,\dots, N-1$ on the Hilbert space. 
The resulting ground state manifold has four states coming from the free edge spins. 
On the ground state manifold, the $\Z_2$ symmetry operator $V$ looks
\begin{align}
P_\theta VP_\theta 
= P(\prod_{j=1}^N \sigma^x_j)P
= \sigma^z_1(\frac{\theta}{2}) \sigma^z_N(\frac{\theta}{2}), 
\end{align}
where $P_\theta$ is the projection onto the ground state manifold. 
Here, the effective $\Z_2$ action on the edge $\sigma^z_1(\frac{\theta}{2}) = \sigma^z_1 e^{\frac{i\theta}{2} \sigma^x_1}$ is the same form as (\ref{eq:edge_Z2_2}). 
Therefore, as discussed in Sec.~\ref{sec:1D_projective_rep_Z2_inv}, the edge $\Z_2$ action has the nontrivial $\Z_2$ invariant of the adiabatic cycle. 

\subsubsection{Kitaev's canonical pump}
Let us consider the model Hamiltonian as well as the ground state of the Kitaev's canonical pump shown in Fig.~\ref{fig:Kitaev_pump} for the $\Z_2$ symmetry operator 
\begin{align}
V = \prod_j \sigma^z_j. 
\end{align}
The ground state shown in Fig.~\ref{fig:Kitaev_pump} is given by 
\begin{align}
\Ket{\Psi_\theta}
=\left\{\begin{array}{ll}
\otimes_j \ket{{\rm I},\theta}_{2j-1,2j} & (\theta\in [0,\pi]), \\
\otimes_j \ket{{\rm II},\theta}_{2j,2j+1} & (\theta\in [\pi,2\pi]), \\
\end{array}\right.
\label{eq:1D_Z2_Kitaev_pump}
 \end{align}
where we have introduced the notations 
\begin{align}
\ket{{\rm I},\theta}_{ij} 
=\cos \frac{\theta}{2} \ket{\ua}_i \ket{\ua}_j
+ \sin \frac{\theta}{2}\ket{\da}_i \ket{\da}_j
\end{align}
and 
\begin{align}
\ket{{\rm II},\theta}_{ij} 
=-\cos \frac{\theta}{2} \ket{\ua}_i \ket{\ua}_j
+ \sin \frac{\theta}{2}\ket{\da}_i \ket{\da}_j. 
\end{align}
A Hamiltonian of which ground state is $\ket{\Psi_\theta}$, which is not unique, is given by the sum of local projection operators 
\begin{align}
H_\theta=
\left\{
\begin{array}{ll}
-\sum_j \ket{{\rm I},\theta}_{2j-1,2j}\bra{{\rm I},\theta}_{2j-1,2j} & (\theta \in [0,\pi]), \\
-\sum_j \ket{{\rm II},\theta}_{2j,2j+1}\bra{{\rm II},\theta}_{2j,2j+1} & (\theta \in [\pi,2\pi]). \\
\end{array}\right. 
\end{align}
It is straightforward to show 
\begin{align}
    H_\theta
    &=-\frac{1}{4}\sum_j \big[
    1+\s^z_{2j-1}\s^z_{2j}+\cos\theta (\s^z_{2j-1}+\s^z_{2j}) \nonumber \\
&    + \sin \theta (\s^x_{2j-1}\s^x_{2j}-\s^y_{2j-1}\s^y_{2j})
    \big]
\end{align}
for $\theta \in [0,\pi]$, and 
\begin{align}
    H_\theta
    &=-\frac{1}{4}\sum_j \big[
    1+\s^z_{2j}\s^z_{2j+1}+\cos \theta (\s^z_{2j}+\s^z_{2j+1}) \nonumber \\
    &- \sin\theta (\s^x_{2j}\s^x_{2j+1}-\s^y_{2j}\s^y_{2j+1})
    \big]
\end{align}
for $\theta \in [\pi,2\pi]$.
The Hamiltonian $H_\theta$ is discontinuous at $\theta=\pi$ but it can be continuous by inserting the following two adiabatic paths of $t \in [0,1]$ at $\theta=\pi$: 
\begin{align}
&-\sum_j \Big\{ (1-t)\frac{1-\sigma^z_{2j-1}}{2} \frac{1-\sigma^z_{2j}}{2} \nonumber \\
&+t (\frac{1-\sigma^z_{2j-1}}{2} + \frac{1-\sigma^z_{2j}}{2} ) \Big\}
\end{align}
and 
\begin{align}
&-\sum_j \Big\{ (1-t) (\frac{1-\sigma^z_{2j-1}}{2} + \frac{1-\sigma^z_{2j}}{2} ) \nonumber \\
&+ t \frac{1-\sigma^z_{2j}}{2} \frac{1-\sigma^z_{2j+1}}{2} 
\Big\}. 
\end{align}
After introducing the matrix product state description of adiabatic cycles in the next section, we see that the state $\Ket{\Psi_\theta}$ is a nontrivial $\Z_2$ cycle. 
See Sec.~\ref{sec:MPS_Kitaev_pump}.

\section{Matrix product states
\label{sec:MPS}}
The discussion in Sec.~\ref{sec:1D_projective_rep_Z2_inv} to define the $\Z_2$ invariant of adiabatic cycles from the edge symmetry action motivates us to formulate the classification of adiabatic cycles by the MPS representation of 1D quantum spin systems.

\subsection{MPS with $\Z_2$ symmetry}
We first generalize the adiabatic cycles in $\Z_2$ symmetric systems by using the MPS.
A translation-invariant MPS is written as 
\begin{align}
    \Ket{\Psi} = 
    \Tr[\cdots A_{m_j} A_{m_{j+1}}\cdots ] \ket{\cdots m_j m_{j+1}.  \cdots } , 
\end{align}
where the index $m_j$ stands for the basis of local Hilbert space at site $j$, and $A_m = [A_m]_{\alpha \beta}$ are $D \times D$ square matrices on the bond Hilbert space.  
A $\Z_2$ symmetry operator is written as a tensor product of a local $\Z_2$ actions 
\begin{align}
    Z = \prod_j \sigma_j, 
\end{align}
where $\sigma_j$ acts on the local Hilbert space as $\sigma_j\ket{m_j} = \ket{n_j} [\sigma_j]_{m_jn_j}$ and satisfies $\s_j^2=1$. 
The uniqueness of the state $\Ket{\Psi}$ is encoded in the matrices $A_m$s:
When $\Ket{\Psi}$ represents a unique gapped ground state, the state $\Ket{\Psi}$ is $\Z_2$ symmetric if and only if there exists a $U(1)$ phase $e^{i\phi}$ and unitary matrix $V \in U(D)$ such that~\cite{PVWC07, PWSVC08}
\begin{align}
    [\sigma]_{mn} A_n = e^{i\phi} V^\dag A_m V
    \label{eq:def_V}
\end{align}
holds. 
We note that the matrix dimension $D$ of $A_m$s reflects the entanglement between two sites. 

\subsubsection{The space of the matrix $V$}
Since the subsequent $\Z_2$ actions and the identity are the same, the uniqueness of the $U(1)$ phase $e^{i\phi}$ and the matrix $V$ guarantees that $e^{i\phi} \in \{\pm 1\}$ and $V$ is a projective representation of $\Z_2$, i.e., $V$ square is proportional to the identity matrix. 
Since $V$ and $e^{i\alpha} V$ represent the equivalent projective representation, the matrix $V$ is regarded as an element of the projective unitary group $PU(D)=U(D)/\{e^{i\alpha} {\bf 1}_D|e^{i\alpha} \in U(1)\}$. 
The constraint on $V$ means that $V^2$ is the identity in the projective unitary group $PU(D)$. 

We are interested in the topological nature of the ``space of gapped $1$D spin systems", especially in the homotopy equivalence class of maps from $S^1$ to that space. 
In the view of MPS representation, the topology of gapped $1$D spin systems may be encoded in the space in which the matrices $A_m$, the factor $e^{i\phi}$, and $V$ live. 
We focus on the space of matrices $V$ and show that its fundamental group is nontrivially given as $\Z_2$.

For the cases of $D=1$ (a trivial tensor product state), the projective unitary group is trivial $PU(1) = \{1\}$, which implies no nontrivial adiabatic cycles. 

For the cases of $D=2$, the projective unitary group is identified with the group of $SO(3)$ rotations $PU(2) \cong SU(2)/\Z_2 \cong SO(3)$. 
Let us write the equivalence class of the matrix $V$ by $[V] = \{z V | z \in U(1)\}$. 
The constraint $V^2 \sim {\bf 1}_2$ implies that $[V]$ is either the identity $[V] = {\rm id}$ of the $SO(3)$ group or a $\pi$-rotation along an axis $\hat n \in S^2$. 
For the former case, we can not have a nontrivial cycle since the space to which $[V]$ belongs is just a point $\{{\rm id}\} \subset SO(3)$. 
On the one hand, we have a nontrivial loop for the latter case. 
Remarkably, $\hat n$ and $-\hat n$ represent the same $\pi$-rotation, $[V]$ belongs to the real projective plane $RP^2 = S^2/\Z_2 \subset SO(3)$ where antipodal points are identified in the 2-sphere $S^2$. 
Therefore, we have a nontrivial loop $\pi_1(RP^2)=\Z_2$ of the space of $[V]$.

To evaluate the fundamental group for generic matrix dimension $D$, let us diagonalize the unitary matrix $V$. 
Due to the constraint $V^2 \sim {\bf 1}_D$, $V$ can be written as 
\begin{align}
    V = z U \left[\begin{array}{cc}
        {\bf 1}_N &  \\
         & -{\bf 1}_{D-N}
    \end{array}\right] U^\dag
\end{align}
with $U$ a $U(D)$ matrix and $z$ a $U(1)$ phase. 
$N$ can be chosen as $0\leq N \leq D/2$, because $z \mapsto -z$ exchanges the eigenvalues $1$ and $-1$. 
Since the multiplication in the form 
\begin{align}
U \mapsto U 
\left[
\begin{array}{cc}
    W &  \\
     & W'
\end{array}
\right]
\label{eq:redef_U}
\end{align}
with $W \in U(N)$ and $W' \in U(D-N)$ does not affect the matrix $V$, the equivalence class $[V]$ belongs to the complex Grassmaniann manifold $G_N(\C^D) = U(D)/(U(N) \times U(D-N))$. 
Since the complex Grassmaniann manifold have the trivial fundamental group $\pi_1(Gr_N(\C^D))=0$, there is no nontrivial adiabatic cycles. 
However, this conclusion is not true when $D = 2N$. 
When $D=2N$, there is an additional identification of matrices $U$ 
\begin{align}
    U \mapsto U \left[\begin{array}{cc}
         & {\bf 1}_N \\
        {\bf 1}_N & 
    \end{array}\right], 
\label{eq:redef_U_2}
\end{align}
which is not a block diagonal matrix in the form \eqref{eq:redef_U}, implying that the equivalence class $[V]$ should be regarded as an element of the quotient space of the Grassmaniann manifold by the $\Z_2$ transformation \eqref{eq:redef_U_2}, i.e., $Gr_N(\C^{2N})/\Z_2$. 
This $\Z_2$ identification is the origin of a nontrivial adiabatic cycle. 
Since $\pi_1(Gr_N(\C^{2N}))=0$, the fundamental group is given by $\pi_1(Gr_N(\C^{2N})/\Z_2) \cong \pi_0(\Z_2) = \Z_2$. 
Therefore, if $D=2N$, a nontrivial adiabatic cycle exists.

The above discussion provides us a simple way to judge if there exists a nontrivial adiabatic cycle: 
There exists a nontrivial adiabatic cycle if and only if the unitary matrix $V$ of projective representation of $\Z_2$ for the bond Hilbert space satisfies $\tr[V]=0$. 

\subsubsection{Gauge invariance of homotopy class}
With the above thought, let us formulate how the $\Z_2$ nontrivial adiabatic cycle is defined from a given cycle of MPS. 
Let $A_m(\theta)$ with $\theta \in[0,2\pi]$ be a cycle of MPS with onsite $\Z_2$ symmetry.
We enforce the periodicity of $A_m(\theta)$. 
Namely, 
\begin{align}
A_m(2\pi) = A_m (0)
\end{align}
for $m=1,\dots, {\rm dim}{\cal H}_j$. 
From onsite $\Z_2$ symmetry, one also has one-parameter families of a $U(1)$ phases $e^{i\phi(\theta)}$ and $U(D)$ matrices $V(\theta)$ by 
\begin{align}
[\sigma_j]_{mn} A_n(\theta)
= 
e^{i\phi(\theta)} V(\theta)^\dag A_m(\theta) V(\theta)
\end{align}
for $\theta \in [0,2\pi]$, where $e^{i\phi(\theta)}$ is unique, and $V(\theta)$ is unique up to a $U(1)$ phase. 
Also, as discussed above, the $\Z_2$-ness ensures that $e^{i\phi(\theta)}$ is a constant $e^{i\phi(\theta)} \equiv \pm 1$, and $V(\theta)^2$ is proportional to the identity matrix. 
Hereafter, we focus on the $U(D)$ matrix $V(\theta)$ only. 
The periodicity of $A_m(\theta)$s implies that the $U(D)$ matrix $V(\theta)$ is also $2\pi$-periodic. 
The equivalence class $[V(\theta)]$ is a loop in the topological space $PU(D)$. 

We should take care about on the gauge choice of $A_m$s. 
The matrices $A_m$ are not unique. 
In fact, the following transformation 
\begin{align}
A_m 
\mapsto \tilde A_m = e^{i \chi} W^\dag A_m W
\label{eq:MPS_gauge_tr}
\end{align}
with $e^{i\chi}$ a $U(1)$ phase and $W$ a $U(D)$ matrix represents the same state as $A_m$. 
However, the homotopy class of $[V(\theta)]$ is found to be independent with this gauge choice.  
Consider a cycle of gauge transformation $(e^{i\chi(\theta)}, W(\theta) )$ with the periodicity $e^{i\chi(2\pi)} = e^{i \chi(0)}$ and $W(2\pi) = W(0)$ so that the gauge transformed matrices $A_m(\theta)$ maintain the periodicity. 
Under the gauge transformation, the $U(D)$ matrix $V(t)$ changes as 
\begin{align}
\tilde V(\theta) = W(\theta)^\dag V(\theta) W(\theta).
\label{eq:V_gauge_tr}
\end{align} 
Since the $U(1)$ phase of $W(\theta)$ does not matter, one can think of $W(\theta)$ as a cycle in the spacial unitary group $SU(D)$ which is contractible, meaning that the cycle $W(\theta)$ is homotopically equivalent to the identity. 
Therefore, $\tilde V(\theta)$ and $V(\theta)$ have the same homotopy class. 

\subsubsection{$\Z_2$ invariant}
\label{sec:MPS_Z2_inv}
The $\Z_2$ invariant for MPSs is defined in the completely same manner as in Sec.~\ref{sec:1D_projective_rep_Z2_inv}. 
Let $\omega(\theta)\in U(1)$ be the two-cocyle (factor system) defined by $V(\theta)^2=\omega(\theta) {\bf 1}$. 
One can define the $\Z_2$ invariant 
\begin{align}
\nu = \frac{1}{2\pi i} \oint d \log \omega(\theta) \quad {\rm mod\ }2. 
\end{align}
as in (\ref{eq:Z2inv}). 
The $\Z_2$-nature is because the redefinition of $U(1)$ phase of $V(\theta)$ changes $\nu$ by an even integer. 

\subsection{Examples}
To illustrate our strategy, we discuss a few examples. 

\subsubsection{MPS of the toy model (\ref{eq:1DZ2_GS})}
We consider the state (\ref{eq:1DZ2_GS}) with a slight modification by a real parameter $r>0$ as 
\begin{align}
    \Ket{\Psi_\theta} 
    =
    \sum_{\{\s_j\}} (re^{\frac{i\theta}{2}})^{N_{\rm dw}} \ket{\cdots \s_j \s_{j+1} \cdots}.
    \label{eq:1DZ2_GS_r}
\end{align}
Namely, the complex wight $r e^{i\theta/2}$ is assigned to each domain wall. 
The MPS for (\ref{eq:1DZ2_GS_r}) reads as 
\begin{align}
A_\ua(\theta) = \begin{pmatrix}
1&r e^{\frac{i\theta}{2}}\\
0&0\\
\end{pmatrix},\quad 
A_\da(\theta) = \begin{pmatrix}
0&0\\
r e^{\frac{i\theta}{2}}&1\\
\end{pmatrix}
\label{eq:1D_MPS_ex_1}
\end{align}
In this gauge choice, $\Z_2$ symmetry is written as 
\begin{align}
[\s^x]_{\s\s'} A_{\s'}(\theta) = \tau^x A_\s(\theta) \tau^x, 
\end{align}
where we have introduced the Pauli matrices $\tau^\mu$ for the bond-Hilbert space.
The gauge choice in (\ref{eq:1D_MPS_ex_1}) breaks the $2\pi$-periodicity, however, it can be $2\pi$-periodic by the gauge transformation 
\begin{align}
A_\s(\theta) \mapsto \tilde A_{\s}(\theta)=W(\theta) A_\s(\theta) W(\theta)^{-1}
\end{align}
with, for example, 
\begin{align}
W(\theta) = \begin{pmatrix}
1\\
&e^{i\theta/2}\\
\end{pmatrix}. 
\end{align}
In doing so, we have a $2\pi$-periodic one 
\begin{align}
\tilde A_\ua(\theta) = \begin{pmatrix}
1&r\\
0&0\\
\end{pmatrix},\quad 
\tilde A_\da(\theta) = \begin{pmatrix}
0&0\\
r e^{i\theta}&1\\
\end{pmatrix}. 
\end{align}
$\Z_2$ symmetry is rewritten as 
\begin{align}
&[\s^x]_{\s\s'}\tilde A_{\s'}(\theta)
=
V(\theta)^\dag\tilde A_\s(\theta)V(\theta),\\
&V(\theta)\sim W(\theta) \tau^x W(\theta)^{-1} = 
\begin{pmatrix}
0&e^{-i\theta/2}\\
e^{i\theta/2}&0\\
\end{pmatrix}. 
\end{align}
This $V(\theta)$ does not satisfy the $2\pi$-periodicity, but since the $U(1)$ phase of $V(\theta)$ is arbitrary, $V(\theta)$ can be $2\pi$-periodic by, for example, 
\begin{align}
V(\theta) = 
\begin{pmatrix}
0&1\\
e^{i\theta}&0\\
\end{pmatrix}. 
\end{align}
We have the nontrivial $\Z_2$ invariant $\nu \equiv 1$ and conclude that the cycle of the MPS (\ref{eq:1D_MPS_ex_1}) belongs to the nontrivial homotopy class. 

One can directly see the matrix $V(\theta)$ wraps a nontrivial $\Z_2$ loop in the topological space $PU(2) \cong SO(3)$. 
The matrix $V(\theta) \sim \cos \frac{\theta}{2} \tau_x + \sin \frac{\theta}{2} \tau_y$ represents the $SO(3)$ $\pi$-rotation around the $(\cos \theta,\sin \theta,0)$-axis. 
Since the $\pi$-rotation around the $(1,0,0)$ and $(-1,0,0)$ are the same, the equivalence class $[V(\theta)]$ forms a nontrivial $\Z_2$ loop.

\subsubsection{MPS of Kitaev's canonical pump}
\label{sec:MPS_Kitaev_pump}
The Kitaev's canonical pump (\ref{eq:1D_Z2_Kitaev_pump}) is invariant under the translation $j \mapsto j+2$. 
Regarding $2j-1$ and $2j$ sites as one site, the matrix product state is given by 
\begin{align}
    &A^{\rm I}_{\ua\ua}(\theta)=\cos \frac{\theta}{2}, \\
    &A^{\rm I}_{\da\da}(\phi)=\sin \frac{\theta}{2}, \\
    &A^{\rm I}_{\ua\da}=A^{\rm I}_{\da\ua}=0, 
\end{align}
for $\theta \in [0,\pi]$, and 
\begin{align}
  &  A^{\rm II}_{\ua\ua}(\theta)=\begin{pmatrix}
    -\cos \frac{\theta}{2}&0\\
    0&0\\
    \end{pmatrix}, \\
   & A^{\rm II}_{\da\da}(\theta)=\begin{pmatrix}
    0&0\\
    0&\sin\frac{\theta}{2}\\
    \end{pmatrix}, \\
&    A^{\rm II}_{\ua\da}(\theta)=\begin{pmatrix}
    0&\sqrt{-\cos\frac{\theta}{2}\sin\frac{\theta}{2}}\\
    0&0\\
    \end{pmatrix}, \\
    &A^{\rm II}_{\da\ua}(\theta)=\begin{pmatrix}
    0&0\\
    \sqrt{-\cos\frac{\theta}{2}\sin\frac{\theta}{2}}&0\\
    \end{pmatrix}, 
\end{align}
for $\theta \in [\pi,2\pi]$. 
The matrix dimensions of $A^{\rm I}(\theta)$ and $A^{\rm II}(\theta)$ are not continuous at $\theta=\pi$. 
To discuss the homotopy class of the MPS, we enlarge the matrix dimension by 2 for $A^{\rm I}(\theta)$ and take the unitary transformation 
\begin{align}
    &A^{\rm I}_{\ua\ua}(\theta)=e^{i \theta\tau_x/2} \begin{pmatrix}
    \cos \frac{\theta}{2}&0\\
    0&0\\
    \end{pmatrix}e^{-i\theta\tau_x/2}, \\
    &A^{\rm I}_{\da\da}(\theta)=e^{i\theta\tau_x/2}\begin{pmatrix}
    \sin \frac{\theta}{2}&0\\
    0&0\\
    \end{pmatrix}e^{-i\theta\tau_x/2}, \\
    &A^{\rm I}_{\ua\da}=A^{\rm I}_{\da\ua}=\begin{pmatrix}
    0&0\\
    0&0\\
    \end{pmatrix}, 
\end{align}
to make the matrices $A^{\rm I}(\theta)$ and $A^{\rm II}(\theta)$ continuous in total. 
We note that the set of matrices $A^{\rm I}_{\s_1\s_2}(\theta)$ have ambiguity as the matrices $A^{\rm I}_{\s_1\s_2}(\theta)$ for $\s_1,\s_2 \in \{\ua,\da\}$ does not generate the algebra of 2 by 2 matrices. 
In fact, $A^{\rm I}_{\s_1\s_2}(\theta)$ commutes with the matrix $e^{i\theta \tau_x/2} e^{i\alpha \tau_z} e^{-i \theta\tau_x/2}$ for any $\alpha$. 
This implies that a unitary matrix defined by (\ref{eq:def_V}) is not unique: 
$V(\theta) \mapsto V(\theta) e^{i\theta\tau_x/2}e^{i\alpha \tau_z}e^{-i\theta\tau_x/2}$ for an arbitrary $\alpha$ satisfies (\ref{eq:def_V}).  

Nevertheless, one can conclude that the adiabatic cycle (\ref{eq:1D_Z2_Kitaev_pump}) belongs to a nontrivial homotopy class. 
The matrix $V(\theta)$ defined by (\ref{eq:def_V}) reads as 
\begin{align}
 V(\theta)\sim 
 \left\{ \begin{array}{ll}
e^{i\theta\tau_x/2}e^{i\alpha(\theta) \tau_z}e^{-i\theta\tau_x/2} & (\theta \in [0,\pi]), \\
i \tau_z & (\theta \in [\pi,2\pi]), 
 \end{array} \right.
\end{align}
where $\alpha(\theta)$ is a real function. 
$V(\theta)$ for $\theta \in [0,\pi]$ is not unique as $\alpha(\theta)$ varies, however, to have a continuous unitary $V(\theta)$, $\alpha(\theta)$ obeys the constraint $\alpha(0), \alpha(\pi) \in \{\pi/2,-\pi/2\}$. 
Note that $V(\theta)^2 \sim {\bf 1}$ does not holds in general. 
Therefore, $V(\theta)$ represents a loop in generic $SO(3)$ rotations, not restricted in $\pi$-rotations. 
Recall that for an $SU(2)$ matrix $V$, the 2 to 1 projection $SU(2) \ni V\mapsto (\bm{n},\varphi) \in SO(3)$, the $\varphi$-rotation around the $\bm{n}$-axis, is given by $\cos \frac{\varphi}{2} = \frac{1}{2} \tr[V]$ and $\sin \frac{\varphi}{2}\bm{n} = \frac{1}{2} \tr[-i \bm{\sigma} V]$. 
The $SO(3)$ parameter of $V(\theta)$ is then extracted as 
\begin{align}
(\bm{n},\phi) =
\left\{ \begin{array}{ll}
\big((0,\sin \theta,\cos \theta),2 \alpha(\theta)\big) & (\theta \in [0,\pi]), \\
\big((0,0,1),\pi\big) & (\theta \in [\pi,2\pi]). \\
\end{array}\right.
\end{align}
Irrespective to the choice of the function $\alpha(\theta)$, $V(\theta)$ wraps a nontrivial loop of the manifold $SO(3)$.

\subsection{Generic finite group symmetry}
\label{sec:MPS_genetic_G}
We generalize the above discussions to generic finite group symmetry which can include antiunitary elements. 
Let $G$ be a finite group and $s: G \to \Z_2=\{1,-1\}$ be the homomorphism specifying if $g \in G$ is unitary $(s_g=1)$ or antiunitary $(s_g=-1)$. 
Let $U(1)_s$ be the $G$-module with the left action $g.z = z^{s_g}$ for $g\in G$ and $z \in U(1)$. 

The following proof is much inspired by \cite{EU16}, where the homotopy type of the space of homomorphisms from a group to the projective unitary group on an infinite dimensional Hilbert space is discussed. 

A simple and translation invariant MPS is written as 
\begin{align}
\ket{\Psi} = \Tr [\cdots A_{m_j} A_{m_{j+1}} \cdots] \ket{\cdots m_j m_{j+1}\cdots}
\end{align}
with $A_m$s a set of $D \times D$ matrices which generates the algebra of $D \times D $ complex matrices and $\ket{m_j}$ the basis of local physical Hilbert space at site $j$. 
The symmetry group $G$ acts on the physical Hilbert space as the tensor product of local actions $\hat g = \bigotimes_j \hat g_j$ with $\hat g_j \ket{m_j} = \ket{n_j} g_{m_jn_j}$. 
Unless misunderstanding arises, we omit the site index $j$. 
At each site, $\hat g_j$ is a linear representation of $G$. 

A simple MPS $\ket{\Psi}$ is invariant under the symmetry group $G$ if and only if there exists a $U(1)$ phase $e^{i\phi_g}$ and a unitary matrix $V_g \in U(D)$ such that $g_{mn} A_n = e^{i\theta_g} V_g^\dag A_m V_g$ for $g \in G$. 
It is found that the $U(1)$ phase $e^{i\theta_g}$ is unique, and the unitary $V_g$ is unique up to a $U(1)$ phase. 
The uniqueness of the set of unitary matrices $V_g$ implies that $V_g$s form a projective representation of $G$, i.e., there exists a two-cocycle $\omega \in Z^2(G,U(1)_s)$ such that $V_gV_h^{s_g}=\omega_{g,h}V_{gh}$ holds. 
Here, we introduced a notation: $V^{s_g} = V$ for $s_g=1$ and $V^{s_g}=V_g^*$ for $s_g=-1$, where $V_g^*$ is the complex conjugation of $V_g$. 
The redefinition $V_g \mapsto \alpha_g V_g$ with $\alpha_g \in C^1(G,U(1)_s)$ induces the change of the two-cocycle $\omega_{g,h} \mapsto \omega_{g,h} \alpha_h^{s_g} \alpha_{gh} \alpha_g$. 
Since the $U(1)$ phase of $V_g$ has no physical meaning, the group cohomology $[\omega] \in H^2(G,U(1)_s) = Z^2(G,U(1)_s)/B^2(G,U(1)_s)$ is regarded as a physical quantity to specify a class of MPS with symmetry $G$~\cite{PETO12,CGW11,SPC11}. 

Let us fix a two-cocycle $\omega \in Z^2(G,U(1)_s)$, and we focus on the space of $\omega$-projective representations themselves. 
There may be additional identification among different $\omega$-projective representations, which comes from a redefinition of the set of $U(1)$ phases of $V_g$s. 
Given a homomorphism $\eta_g \in \Hom(G,U(1)_s) = H^1(G,U(1)_s)$, which satisfies $\eta_g\eta_h^{s_g}=\eta_{gh}$, the redefining of $V_g$ by $V_g \mapsto \eta_g V_g$ may or may not change the $\omega$-projective representation $V_g$ while keeping the two-cocycle $\omega$.

Let $\rho_1,\rho_2,\dots$ be the equivalence classes of irreducible $\omega$-projective representations. 
The equivalence class of an $\omega$-projective representation $V$ is a direct sum $V \sim \bigoplus_a \rho_a^{\oplus n_a}$ of $\rho_a$s with $n_a$s nonnegative integers representing the number of $\rho_a$ irreps in $V$. 
Let $X^\omega_{\vec{n}}$ be the space of $\omega$-projective representations of which the equivalence class is $\bigoplus_a \rho_a^{\oplus n_a}$. 
Here we introduced a vector notation $\vec{n}=(n_1,n_2,\dots)$. 
The total space $X^\omega$ of $\omega$-projective representations is the disjoint union $X^\omega =\amalg_{\vec{n}} X^\omega_{\vec{n}}$.
The group $\Hom(G,U(1)_s)$ acts on the total space $X^\omega$ by $(\eta V)_g = \eta_g V_g$. 
Since $V_g$ and $\eta_g V_g$ with $\eta \in \Hom(G,U(1)_s)$ are regarded as physically the same action, the space of symmetry action on the bond Hilbert space can be identified with the quotient space $X^\omega\big/\Hom(G,U(1)_s)$.  
Therefore, the adiabatic cycles of the MPS with $G$ symmetry is classified by the homotopy equivalence class 
\begin{align}
\Big[S^1,X^\omega\Big/\Hom(G,U(1)_s)\Big]. 
\end{align}

Let us focus on an orbit 
\begin{align}
    \bigcup_{\eta \in \Hom(G,U(1)_s)} X^\omega_{\eta(\vec{n})}
\end{align}
to which a given $\omega$-projective representation $V_g$ with the vector $\vec n$ belongs. 
We denote the dimension of $V_g$ by $D$. 
The quotient space is given by 
\begin{align}
    &\bigcup_{\eta \in \Hom(G,U(1)_s)} X^\omega_{\eta(\vec{n})}
    \Big/ \Hom(G,U(1)_s) \nonumber\\
    &\cong 
    X^\omega_{\vec{n}}\Big/\Hom(G,U(1)_s)_{\vec{n}}, 
\end{align}
where we have introduced the stabilizer subgroup $\Hom(G,U(1))_{\vec{n}}:=\{\eta \in \Hom(G,U(1)_s) | \eta(\vec{n})=\vec{n}\}$. 
Elements of the stabilizer subgroup $\Hom(G,U(1))_{\vec{n}}$ represent the homomorphisms $\eta \in \Hom(G,U(1))$ that does not change the equivalence class of the $\omega$-representation specified by $\vec n$. 
We find that the space $X^\omega_{\vec{n}}$ is simply connected: 
Every representation $V_g$ belonging to the equivalence class $\vec{n}$ is written as $V_g = W V_g^{\rm ref} W^\dag$ with $V^{\rm ref}_g$ a reference representation and $W$ a unitary matrix $W \in U(D)$. 
Since the $U(1)$ phase part of $W$ does not change $V_g$, $W$ can be an element of the spacial unitary group $SU(D)$ that is simply connected. 
Then, a loop $V(\theta): S^1 \to X^\omega_{\vec{n}}$ of $\omega$-projective representations can be written as  $V_g(\theta) = W(\theta)V^{\rm ref}_g W(\theta)^\dag$ with $W: S^1 \to SU(D)$ a loop on $SU(D)$. 
Since $SU(D)$ is simply connected, there is a homotopy equivalence $W(\theta) \sim {\bf 1}_D$, which gives the homotopy equivalence of $V(\theta)$, $V_g(\theta) \sim V_g^{\rm ref}$. 
Thus, we conclude that 
\begin{align}
   & \Big[S^1, X^\omega_{\vec{n}}\Big/\Hom(G,U(1)_s)_{\vec{n}} \Big] \nonumber \\
    & \cong \Hom(G,U(1)_s)_{\vec{n}}. 
\end{align}
This is the central result of this section. 
The adiabatic cycles of the MPS is classified by the stabilizer subgroup $\Hom(G,U(1)_s)_{\vec n}$, which is the space of $G$ symmetry charges keeping the $\omega$-projective representation invariant as an equivalence class. 


There is a practical method to calculate the stabilizer subgroup $\Hom(G,U(1)_s)_{\vec{n}}$ for a given projective representation $V_g$. 
Firstly, it is sufficient to consider the center group $Z(G_0) = \{g \in G_0 | gh=hg {\rm\ for\ all\ }h \in G_0\}$ of the unitary subgroup $G_0 = \ker (s) = \{g \in G | s_g=1\}$. 
If there exists $g \in Z(G_0)$ such that the both $\eta_g\neq 1$ and $\tr[V_g]\neq 0$ hold, the representation $\eta V$ defined by $(\eta V)_g = \eta_g V_g$ is not equivalent to $V$, because of the mismatch of the $\omega$-projective character $\tr[V_g]$. 
The converse is also true. 
Thus, we arrive at the following statement. 
For an MPS with a projective representation $V_g$ of $G$, the classification of adiabatic cycle is given by the subgroup of $\Hom(G,U(1)_s)$ composed of the elements $\eta \in \Hom(G,U(1)_s)$ such that $\tr[V_g] = 0$ holds for all $g \in Z(G_0)$ with $\eta_g\neq 1$.

\subsection{Topological invariant from two-cocycle}
Let $V_g(\theta)$ be the $G$-action on the bond Hilbert space of the MPS $A_m(\theta)$ defined by $g_{mn} A_n(\theta) = e^{i \phi_g(\theta)} V_g(\theta)^\dag A_m(\theta) V_g(\theta)$. 
$V_g(\theta)$ is a projective representation of $G$ with a two-cocycle $\omega_{g,h}(\theta) \in Z^2(G,U(1)_s)$ which also depends on $\theta$. 
Namely, $V_g(\theta) V_h(\theta)^{s_g}= \omega_{g,h}(\theta) V_{gh}(\theta)$. 
$V_g(\theta)$s can be chosen to be $2\pi$-periodic. 
In doing so, the two-cocycle $\omega_{g,h}(\theta)$ is also $2\pi$-periodic, and one can define the $\Z$-valued winding number
\begin{align}
    n_{g,h}= \frac{1}{2 \pi i} \oint d \log \omega_{g,h}(\theta) \in \Z
    \label{eq:1D_MPS_top_inv}
\end{align}
for each pair $(g,h)$. 
The cocycle condition of $\omega_{g,h}(\theta)$ implies that $n_{g,h}$ is a two-cocycle of $Z^2(G,\Z_s)$, where $\Z_s$ is the $G$-module with the left-$G$-action $g \cdot n = s_g n$ for $n \in \Z$.  
A redefinition $V_g(\theta) \mapsto V_g(\theta) \alpha_g(\theta)$ with a $2\pi$-periodic $U(1)$-valued functions $\alpha_g(\theta)$ changes the two-cocycle by the two-coboundary $(d\alpha(\theta))_{g,h}=\alpha_h(\theta)^{s_g}\alpha_{gh}(\theta)^{-1}\alpha_g(\theta)$. 
The two-coboundary $d \alpha(\theta)$ defines the set of winding numbers by $(d m)_{g,h}=s_g m_h - m_{gh} + m_g \in B^2(G,\Z_s)$ with 
\begin{align}
    m_g = \frac{1}{2\pi i} \oint d \log \alpha_g(\theta) \in \Z
\end{align}
for $g \in G$. 
This gives the equivalence relation of two-cocycles $Z^2(G,\Z_s)$. 
We conclude that the topological invariant of adiabatic cycles of MPSs lives in the cohomology group 
\begin{align}
[n]\in H^2(G,\Z_s) = Z^2(G,\Z_s)/B^2(G,\Z_s). 
\end{align}

The isomorphism $H^2(G,\Z_s) \cong H^1(G,U(1)_s)$ suggests that the invariant (\ref{eq:1D_MPS_top_inv}) can be interpreted as the pumped charge of the symmetry group $G$ by a period. 
As shown in Sec.~\ref{sec:any_dim}, we can indeed construct a model of 1D adiabatic cycle from a given element of $Z^1(G,U(1)_s)$ by using the Bockstein homomorphism. 

If the cohomology group $H^2(G,U(1)_s)$ is not trivial, the two-cocycle $\omega_{g,h}(\theta)$ can run over a nontrivial sector of $Z^2(G,U(1)_s)$. 
This means the set $[S^1,Z^2(G,U(1))]$ of homotopy equivalence classes of map $S^1 \to Z^2(G,U(1))$ splits into the sectors by $H^2(G,U(1)_s)$. 
For each sector, one can define the winding number $n_{g,h}$ in the same way. 
Thus, the homotopy equivalence class is classified by 
\begin{align}
\frac{[S^1,Z^2(G,U(1)_s)]}{[S^1,B^2(G,U(1)_s)]} 
&\cong H^2(G,U(1)_s) \times H^2(G,\Z_s).
\end{align}
This is in complete agreement with the classification of Floquet SPTs in 1D.~\cite{EN16,PMV16,KS16}

\section{A two-dimensional model of adiabatic pump with time-reversal symmetry}
\label{sec:2DTRS}

In this section we present an exactly solvable model of the adiabatic pump in 2-spatial dimensions with time-reversal symmetry (TRS). 

\subsection{Model}
\label{sec:2D_model}
We consider a model slightly modified from Levin--Gu model~\cite{LG12}, which is a prototypical model for SPT phases in 2D. 
In the same way as in Sec.~\ref{sec:1D_model}, we start up with the trivial paramagnet as the model for the initial parameter, and take a local unitary transformation with $\theta$. 
Let us consider the spin 1/2 degrees of freedom on the triangular lattice. 
We denote the spin operator at site $j$ by $\sigma^\mu_j$ for $\mu=x,y,z$. 
The initial Hamiltonian is 
\begin{align}
H_0 = -\sum_j \s^x_j
\end{align}
We apply the local unitary~\cite{LG12,SF16}
\begin{align}
U_\theta 
&= \prod_{<pqr>} e^{\frac{i\theta}{24}(3\s^z_p\s^z_q\s^z_r-\s^z_p-\s^z_q-\s^z_r)} \nonumber \\
&= \prod_j e^{-\frac{i\theta}{12} \s^z_j \sum^j_{pq}\frac{1-\s^z_p\s^z_q}{2}}
\label{eq:2D_TRS_local_unitary}
\end{align}
to $H_0$. 
Here,  $<pqr>$ runs over all triangles, and the sum $\sum^j_{pq}$ means that $pq$ stands for all the nearest neighbor links of $j$. 
Here we showed two expressions in (\ref{eq:2D_TRS_local_unitary}), the same local unitary in bulk but different with a boundary. 
We define the adiabatic Hamiltonian by 
\begin{align}
H_\theta = U_\theta H_0 U_\theta^{-1} 
= - \sum_j B_j^\theta, 
\label{eq:2D_model}
\end{align}
with 
\begin{align}
B_j^\theta 
= U_\theta \s^x_j U_\theta^{-1} 
= \s^x_j e^{\frac{i\theta}{2} \s^z_j \sum^j_{pq} \frac{1-\s^z_p\s^z_q}{2}} e^{-i\theta \s^z_j}
\end{align}
We find that $B_j^{2\pi}=\sigma^x_j$, meaning that the periodicity of $H_\theta$ is $2\pi$. 
In particular, $H_{\theta=\pi}$ is recast as the Levin--Gu model as $H_{\theta=\pi}$ has $\Z_2$ symmetry defined by $\prod_j \sigma^x_j$. 
For generic $\theta$, no unitary $\Z_2$ symmetry exists, but there is TRS defined by 
\begin{align}
T=(\prod_j \s^x_j) {\cal K}, 
\end{align}
where we have denoted the complex conjugation by ${\cal K}$.

\begin{figure}[!]
\centering
\includegraphics[width=\linewidth, trim=0cm 0cm 0cm 0cm]{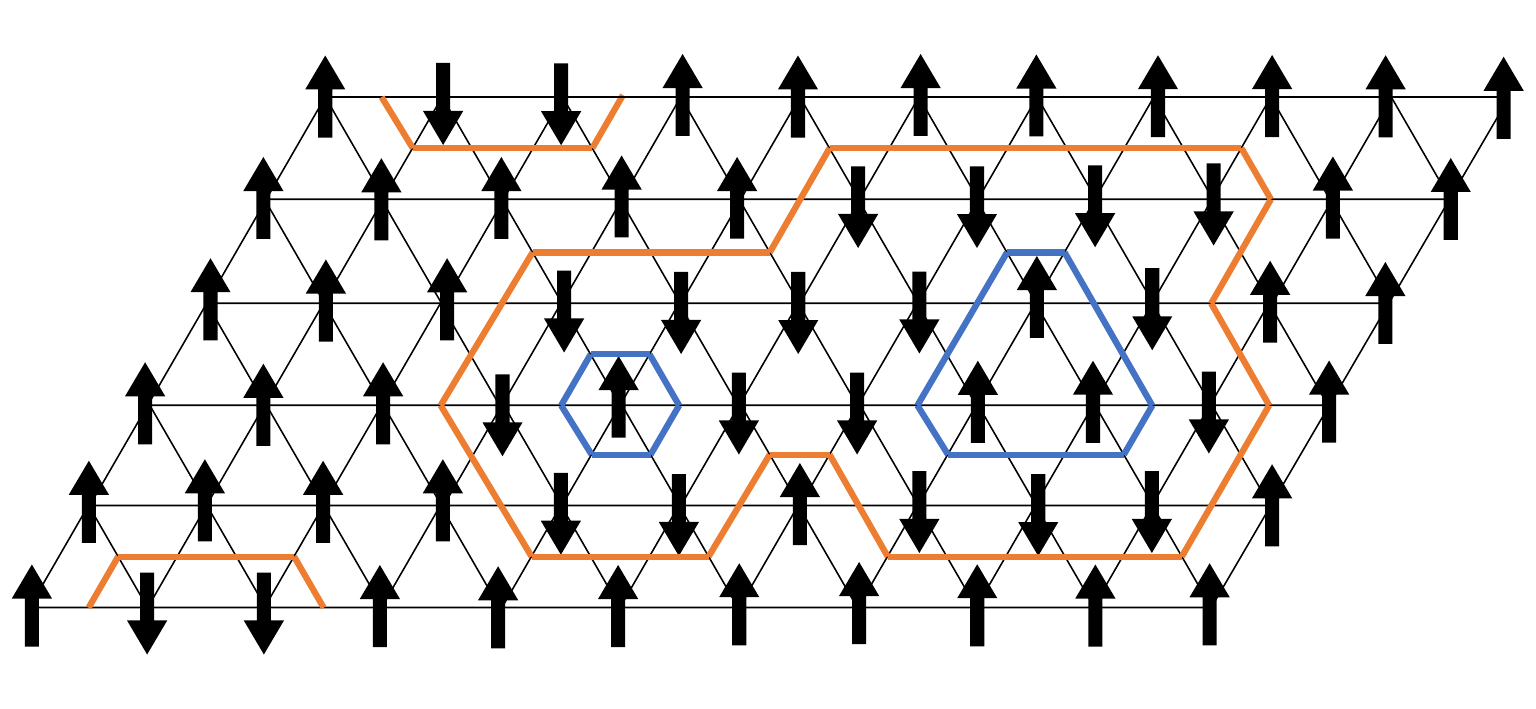}
\caption{The wave function (\ref{eq:2D_wave_func}) over a torus. 
This figure shows a spin configuration with $N_\ua=N_\da=2$. 
The $U(1)$ phase factor $e^{i\theta}$ ($e^{-i\theta}$) is attached to the blue (orange) loops. }
\label{fig:model_2d}
\end{figure}

On a closed manifold, the ground state is unique, as is $H_0$, and the ground state wave function is given by 
\begin{align}
\braket{\{\s_j\} | \Psi} = e^{i\theta (N_\ua-N_{\da})} 
\label{eq:2D_wave_func}
\end{align}
on the basis of $\sigma^z_j =\pm 1$. 
Here, $N_\ua$ ($N_\da$) is the number of contractible loops whose interior near the loop is up (down) spins. 
See Fig.~\ref{fig:model_2d} for a snapshot wave function. 
It should be noted that no $U(1)$ phases are attached to the non-contractible loops. 

On a closed manifold, the local unitary $U_\theta$ is $2\pi$-periodic and preserves TRS. 
However, on an open manifold like an open disk, depending on local terms near the boundary, $U_\theta$ can be either $2\pi$-periodic or time-reversal symmetric. 
This issue is discussed from a more general perspective in Sec.~\ref{eq:any_dim_local_unitary}.

\subsection{Open disk}
We consider the model (\ref{eq:2D_model}) on an open disk. 
The calculations in this section is almost parallel to Ref.~\cite{LG12}. 
The Hamiltonian is of the form $H_\theta = H_\theta^{\rm bulk} + H_\theta^{\rm bdy}$ with $H_\theta^{\rm bulk}= \sum_{j \in {\rm bulk}}B^\theta_j$, and $H_\theta^{\rm bdy}$ is composed of local Hamiltonians near the boundary with $2\pi$-periodicity and TRS. 
Here, the sum $\sum_{j \in {\rm bulk}}$ runs over sites strictly interior of the system. 
We first solve $H_\theta^{\rm bulk}$ to get the degenerate ground state manifold and discuss the effect of the boundary Hamiltonian $H^{\rm bdy}_\theta$ from the degenerate perturbation theory. 

We denote the site index on the boundary by $n \in {\rm bdy}$. 
The ground state manifold of $H^{\rm bulk}_\theta$ is specified by the boundary spins $\sigma_n \in \pm 1$ as 
\begin{align}
&\Ket{\Psi_\theta(\{\sigma_{n \in {\rm bdy}}\})} \nonumber\\
&\sim \prod_{j \in {\rm bulk}} (1+B_j^\theta) 
\Ket{\{\sigma_{j\in {\rm bulk}} \equiv 1\}, \{\sigma_{n\in {\rm bdy}}\}}.
\end{align}
The relative $U(1)$ phases among ground states $\Ket{\Psi_\theta(\{\sigma_n\})}$ are undetermined in general. 
However, as seen in Sec.~\ref{sec:comment_relative_phases}, to make the effective boundary Hamiltonian local one, it is important to satisfy a kind of locality for the choice of the relative phases among $\Ket{\Psi_\theta(\{\sigma_n\})}$. 
Here we employ the same prescription as Ref.~\cite{LG12}. 
We assume the ``ghost spins'' outside of the system and fix these spins to the up states. 
With this prescription, the relative phases are determined as 
\begin{align}
\Braket{\{\sigma_{j \in {\rm bulk}}\}|\Psi_\theta(\{\sigma_n\})}
= 
e^{i\theta (N_\ua-N_\da)}, 
\end{align}
where $N_\ua$ and $N_\da$ are the ones introduced before. 

Introduce the spin operators $\bar \sigma^\mu_n$ with $\mu=x,y,z$ acting on the ground state manifold $\Ket{\Psi_\theta(\{\sigma_n\})}$. 
Note that $\bar \sigma^\mu_n$ is different from $\sigma_n^\mu$, the original spin operators on the boundary. 
Let $P_\theta$ be the projection onto the ground state manifold. 
One can find the TRS operator on the ground state manifold is
\begin{align}
\bar T_\theta
&:= P_\theta T P_\theta \nonumber \\
&=
\prod_{n} (\bar \sigma^x_n e^{i\theta} e^{\frac{i\theta}{2} \frac{1-\bar \sigma^z_n \bar \sigma^z_{n+1}}{2}}) {\cal K}  \nonumber \\
&\sim 
(\prod_{n} \bar \sigma^x_n) (\prod_n e^{\frac{i\theta}{2} \frac{1-\bar \sigma^z_n \bar \sigma^z_{n+1}}{2}}) {\cal K}. 
\label{eq:local_TRS}
\end{align}
Here, we have ignored an unimportant $U(1)$ phase factor. 
The unitary part of $\bar T_\theta$ is not a product of a unitary operator at each site, which is a characteristic feature of SPT phases in 2D.~\cite{LG12,CLW11} 
Note that without the edge of the boundary, $(T_\theta^{\rm bdy})^2 = 1$ holds. 

\subsection{Microscopic edge theory}

\begin{figure}[!]
\centering
\includegraphics[width=\linewidth, trim=0cm 0cm 0cm 0cm]{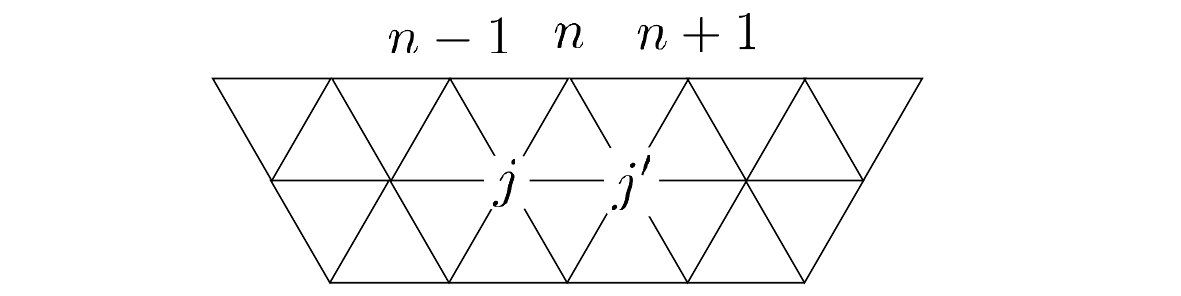}
\caption{Labeling sites near the boundary.}
\label{fig:2d_bdy}
\end{figure}

Following Ref.~\cite{LG12}, we first introduce the boundary local Hamiltonian $B^{\ua,\theta}_n$ to be the same form as bulk ones $B_j^\theta$ but with the fixed ghost spins outside of the system 
\begin{align}
&B^{\ua,\theta}_n \nonumber \\
&=
\s^x_n e^{\frac{i\theta}{2} \s^z_n (\frac{1-\s^z_{n-1}}{2} + \frac{1-\s^z_{n+1}}{2} + \sum^n_{<jj'>} \frac{1-\s^z_j \s^z_{j'}}{2})} e^{-i\theta \sigma^z_n}.
\label{eq:2D_TRS_bdy_spin_1}
\end{align}
Here, $\sum^n_{jj'}$ runs over the triangles $<njj'>$ containing the boundary sites $n$ (See Fig.~\ref{fig:2d_bdy}).
The advantage of this boundary term is that $B^{\ua,\theta}_n$ commutes with bulk ones $B_j$, meaning that the eigenstates of the effective edge Hamiltonian is the exact eigenstates of the total system. 
On the ground state manifold, we have 
\begin{align}
P_\theta B^{\ua,\theta}_n P_\theta
=
\bar \s^x_n. 
\end{align}
But this does not satisfy TRS $\bar T_\theta$. 
To enforce TRS, we add the local term 
\begin{align}
\bar T_\theta \bar \s^x_n \bar T_\theta^{-1} 
= 
\bar \s^x_n e^{\frac{i\theta}{2}\bar \s^z_n (\bar \s^z_{n-1}+\bar \s^z_{n+1})}
\end{align}
to get the $2\pi$-periodic and time-reversal symmetric effective boundary Hamiltonian 
\begin{align}
\bar H^{\rm bdy}_\theta
&:=P_\theta H_\theta^{\rm bdy} P_\theta \nonumber \\
&=
-\lambda \sum_{n} (\bar \s^x_n + \bar \s^x_n e^{\frac{i\theta}{2}\bar \s^z_n (\bar \s^z_{n-1}+\bar \s^z_{n+1})})
\label{eq:2D_bdy_hamiltonian}
\end{align}
with $\lambda$ a small constant. 

We would like to prove that the ground states of any one-parameter family of effective boundary Hamiltonians $\bar H_\theta$ respecting TRS $\bar T_\theta$ can not be unique for all $\theta \in [0,2\pi]$. 
In this paper, we could not prove this no-go. 
In the rest of this section, we leave a discussion on ingappability of the effective boundary Hamiltonian of the form (\ref{eq:2D_bdy_hamiltonian}). 

\subsubsection{Discussion: fermionic dual model and ingappability}
The effective Hamiltonian (\ref{eq:2D_bdy_hamiltonian}) accidentally has $\Z_2$ onsite symmetry defined by $\prod_n \sigma^x_n$ in addition to TRS $\bar T_\theta$. 
Applying the Kramers--Wannier duality map (\ref{eq:KW_1},\ref{eq:KW_2}) and the Jordan--Wigner transformation (\ref{eq:JW_1}, \ref{eq:JW_2}), and (\ref{eq:JW_3}) to the effective Hamiltonian (\ref{eq:2D_bdy_hamiltonian}), we get the dual fermion model called the Kitaev chain~\cite{Kit01}
\begin{align}
\check H_\theta
&=
2 \lambda \sum_n [
-a^\dag_n a_{n+1}-a^\dag_{n+1}a_n \nonumber\\
&+e^{\frac{i\theta}{2}} \cos \frac{\theta}{2}a^\dag_na^\dag_{n+1}+e^{-\frac{i\theta}{2}} \cos \frac{\theta}{2} a_{n+1}a_n
]
\label{eq:Kitaev_chain_theta}
\end{align}
and the dual TRS 
\begin{align}
\check T_\theta
=
e^{\frac{i\theta}{2}\sum_n a^\dag_na_n} {\cal K}.
\end{align}
Importantly, the dual TRS is not $2\pi$-periodic, while obeys 
\begin{align}
\check T_{2\pi}=(-1)^F {\cal K}, 
\end{align}
where $(-1)^F$ is the fermion parity operator. 
This is the symmetry class BDI in fermionic SPT phases. 
The classification is known to be $\Z_8$~\cite{FK11}. 

We first discuss ingappability as free fermions, where the topological classification is $\Z$ which is characterized by some winding number. 
The BdG Hamiltonian ${\cal H}^\theta_{\rm BdG}$ defined by $\check H_\theta = \frac{1}{2} \sum_{n,n'} [{\cal H}^\theta_{\rm BdG}]_{n,n'}$ has the $\theta$-dependent chiral symmetry $\Gamma_\theta{\cal H}^\theta_{\rm BdG} \Gamma_\theta^{-1} = - {\cal H}^\theta_{BdG}$ with  $\Gamma_\theta = \begin{pmatrix}
0&e^{i\theta/2}\\
e^{-i\theta/2}&0\\
\end{pmatrix}$. 
With transnational invariance, the winding number $N_w$ is written by the BdG Hamiltonian ${\cal H}_{\rm BdG}^\theta(k)$ in the Bloch-momentum space as $N_w = \frac{1}{4\pi i} \oint dk \tr \big[\Gamma_\theta [{\cal H}_{\rm BdG}^\theta(k)]^{-1} \partial_k  {\cal H}_{\rm BdG}^\theta(k) \big] \in \Z$. 
Since the winding number $N_\omega$ is quantized as it takes a value in integers, $N_\omega$ remains a constant unless no gapless points of ${\cal H}_{\rm BdG}^\theta(k)$ arise for all $\theta \in [0,2\pi]$. 
On the one hand, the periodicity of ${\cal H}_{\rm BdG}^\theta$ and $\Gamma_{2\pi}=-\Gamma_0$ imply that $N_\omega=0$ if no gapless points arise. 
Therefore, if $N_\omega \neq 0$ at $\theta=0$, there must be a gapless point at some $\theta \in (0,2\pi)$. 
For example, the Kitaev chain (\ref{eq:Kitaev_chain_theta}) shows $|N_\omega|=1$ at $\theta=0$, and it is consistent with the gapless point of (\ref{eq:Kitaev_chain_theta}) at $\theta=\pi$. 

For the many-body fermionic Hilbert space, ingappability is more subtle since we have to distinguish $1 \in \Z_8$ and $-1 \in \Z_8$ phases. 
Namely, a $\Z_4$ many-body invariant is needed. 
In the Euclidean space-time path-integral picture, the $\Z_8$ invariant $\nu \in \Z_8$ is known to be the discrete $U(1)$ phase of the partition function over the real projective plane $\R P^2$, $Z(\R P^2,\pm) = |Z(\R P^2,\pm)|e^{\pm \frac{2\pi i \nu}{8}}$~\cite{KTTW15}.  
Here, $\pm$ means two different Pin$_-$-structure on $\R P^2$, which is exchanged by the local fermion parity transformation on the orientation-reversing patch intersection introduced by the TRS operator $\check T_\theta$. 
Then, the relation $\check T_{2\pi}=(-1)^F \check T_0$ implies that the $\Z_8$ invariant should satisfy $\nu \equiv -\nu$ modulo 8 if the ground state is unique. 
This is only consistent when $\nu \equiv 0,4$ modulo 8, implying that if $\nu \equiv 1,2,3$ modulo 4 at $\theta=0$, there must be a phase transition in $\theta \in (0,2\pi)$. 

We note that the discussion in this section is based on the assumption of the additional $Z_2$ symmetry by $\prod_n \sigma^x_n$.

\subsection{$\Z_2$ invariant from three-cocycle}

In this section, we discuss how the adiabatic cycle of (\ref{eq:2D_model}) is nontrivial in the viewpoint of the three-cocycle. 
Since the ground state of (\ref{eq:2D_model}) is unique and symmetric, we can in principle extract the one-parameter family of three-cocycle $\omega_\theta \in Z^3(\Z_2^T,U(1)_s)$ characterizing the ground state with TRS, where $U(1)_s$ the left $\Z_2$-module defined in the same way as in Sec.~\ref{sec:MPS_genetic_G}. 

\subsubsection{$\Z_2$ invariant}
Before computing the three-cocycle of the ground state (\ref{eq:2D_wave_func}), we first investigate how the space $Z^3(\Z_2^T,U(1)_s)$ looks like. 
Solving the cocycle condition 
\begin{align}
&(d \omega)(g,h,k,l) 
= \omega(h,k,l)^{s_g} \omega(gh,k,l)^{-1} \nonumber\\
&\qquad \qquad\omega(g,hk,l) \omega(g,h,kl)^{-1} \omega(g,h,k) = 0
\end{align}
directly, we have $Z^3(\Z_2^T,U(1)_s) \cong U(1)^3$ which is independently parameterized by, for example, $\omega(e,e,T), \omega(T,T,e)$ and $\omega(T,T,T)$. 
Therefore, given a $2\pi$-periodic three-cocycle $\omega_\theta$, one can define three $\Z$ invariants as winding numbers of these representatives. 
However, a part of $\Z$ invariants is trivialized by $2\pi$-periodic three-coboundaries $d\alpha_\theta$ with $\alpha_\theta \in C^2(\Z_2^T,U(1)_s)$. 
Under the three-coboundary $d\alpha$, $\omega$ changes as 
\begin{align}
&\omega(e,e,T) \nonumber\\
&\mapsto \omega(e,e,T)\alpha(e,T)\alpha(e,e)^{-1}, \\
&\omega(T,T,e) \nonumber\\
&\mapsto \omega(T,T,e) \alpha(T,e)^{-1}\alpha(e,e)^{-1}, \\
&\omega(T,T,T)\nonumber\\
&\mapsto \omega(T,T,T)\alpha(T,T)^{-2}\alpha(e,T)^{-1}\alpha(T,e).
\end{align}
Therefore, the only one $\Z_2$ invariant is well-defined. 
Explicitly, given a $2\pi$-periodic three-cocycle $\omega_\theta$, the $\Z_2$ invariant is defined by 
\begin{align}
\nu = \frac{1}{2\pi i} \oint d \log [\omega_\theta(e,e,T)\omega_\theta(T,T,e)\omega_\theta(T,T,T) ]
\label{eq:2D_TRS_Z2_inv}
\end{align}
modulo 2. 

\subsubsection{Review on Else-Nayak's method}
We adapt the method in Ref.~\cite{EN14}, where they showed how the $(d+1)$-cocycle emerges from the local (anti)unitaries defined on the $(d-1)$-dimensional boundary. 
Let $G$ be a symmetry group possibly including antiunitary elements and $U(g \in G)$ be the local symmetry action on the 1D boundary of a 2D non-chiral invertible state. 
Because $U(g)$ is written with local operators, one can restrict $U(g)$ on an interval $I=[a,b]$ to get the symmetry action $U_I(g)$ on the interval $I$. 
$U_I(g)$ is only defined modulo local unitaries acting near the edge of $I$, which leads the breaking the group law near the edge as in  
\begin{align}
U_I(g) U_I(h) = \Omega_{\partial I}(g,h) U_I(gh), 
\end{align}
where $\Omega_{\partial I}(g,h)$ is a local unitary near the edge $\partial I$. 
The associativity of $U_I(g)$s implies that the following constraint condition on $\Omega_{\partial I}(g,h)$, 
\begin{align}
\Omega_{\partial I}(g,h) \Omega_{\partial I}(gh,k)
= {}^{U_I(g)} \Omega_{\partial I}(h,k) \Omega_{\partial I}(g,hk) 
\label{eq:EN_condition}
\end{align}
with ${}^{U_I(g)} \Omega_{\partial I}(h,k) = U_I(g) \Omega_{\partial I}(h,k) U_I(g)^{-1}$. 
We further restrict $\Omega_{\partial I}(g,h)$ to the left edge part, which we denote by $\Omega_a(g,h)$. 
For $\Omega_a(g,h)$, the condition (\ref{eq:EN_condition}) holds true only modulo a $U(1)$ phase 
\begin{align}
&\Omega_a(g,h) \Omega_a(gh,k)\nonumber \\
&=\omega(g,h,k) {}^{U_I(g)} \Omega_a(h,k) \Omega_a(g,hk). 
\label{eq:EN_condition_2}
\end{align}
It is shown that $\omega(g,h,k)$ defined in (\ref{eq:EN_condition_2}) satisfies the three-cocycle condition. 

\subsubsection{Boundary TRS and three-cocycle}
We consider the local antiunitary (\ref{eq:local_TRS}) as the TRS operator on an interval. 
For our purpose to extract the $2\pi$-periodic three-cocycle, the local antiunitary on the interval should also be $2\pi$-periodic. 
Such a $2\pi$-periodic local antiunitary is 
\begin{align}
U^\theta_I(T)=
(\prod_{n=1}^N \sigma^x_n )
(\prod_{n=1}^{N-1} e^{i\theta\frac{1+\sigma^z_n}{2}\frac{1-\sigma^z_{n+1}}{2}} )
{\cal K}
\end{align}
for TRS and $U^\theta_I(e) = {\rm Id}$. 
One can reads off the boundary unitaries $\Omega^\theta_{\partial I}(g,h)$ parameterized by $\theta$ as 
\begin{align}
\Omega^\theta_{\partial I}(T,T) 
&=  e^{-\frac{i\theta}{2}\sigma^z_1} e^{\frac{i\theta}{2}\sigma^z_N}
\end{align}
and $\Omega^\theta_{\partial I}(g,h) = {\rm Id}$ otherwise. 
Restricting $\Omega^\theta_{\partial I}(T,T)$ to the left edge, we have $\Omega^\theta_a(T,T) = e^{i\theta \frac{1-\sigma^z_1}{2}}$ which is $2\pi$-periodic. 
The three-cocycle defined by (\ref{eq:EN_condition_2}) is given by $\omega_\theta(T,T,T) = e^{i\theta}$ and $\omega_\theta(g,h,k)=1$ otherwise, resulting in the nontrivial $\Z_2$ invariant (\ref{eq:2D_TRS_Z2_inv}) $\nu \equiv 1$. 
Thus, the one-parameter family of the local TRS (\ref{eq:local_TRS}) forms a $\Z_2$ nontrivial loop which can not be deformed to a constant local TRS. 

\subsection{Haldane chain pump}
Interestingly, the local unitary (\ref{eq:2D_TRS_local_unitary}) by a period is viewed as pumping a Haldane chain protected by TRS on the boundary of 2D systems~\cite{PM16,RH17,TV21}.
To see this, we consider the local unitary on the open disk in the form 
\begin{align}
U_\theta^{\rm open}
= \prod_{<pqr>} e^{\frac{i\theta}{24}(3\s^z_p\s^z_q\s^z_r-\s^z_p-\s^z_q-\s^z_r)}, 
\end{align}
where $<pqr>$ runs over the all triangles of the open disk. 
Note that $U^{\rm open}_\theta$ is chosen to have TRS, but no $2\pi$-periodicity. 
A spin operator $\sigma^\mu_{n \in {\rm bdy}}$ on the boundary transforms as 
\begin{align}
\tilde B^{\theta,\mu}_n
&=U_\theta^{\rm open} \sigma^\mu_n (U_\theta^{\rm open})^{-1}\nonumber\\
&=\sigma^\mu_n e^{\frac{i\theta}{2} \sigma^z_n \sum^n_{<jj'>} \frac{1-\sigma^z_j \sigma^z_{j'}}{2}} e^{-\frac{i\theta}{2} \sigma^z_n} 
\end{align}
for $\mu=x,y$ and $\tilde B^{\theta,z}_n=U_\theta^{\rm open} \sigma^z_n (U_\theta^{\rm open})^{-1}=\sigma^z_n$. 
Here, the sum $\sum^n_{<jj'>}$ runs over the all triangles containing the boundary site $n$ (See Fig.~\ref{fig:2d_bdy}). 
Note the difference from the boundary interaction (\ref{eq:2D_TRS_bdy_spin_1}) introduced before. 
(\ref{eq:2D_TRS_bdy_spin_1}) preserves the $2\pi$-periodicity, but breaks TRS. 
In contrast, $\tilde B^{\theta,\mu}_n$ preserves TRS but breaks the $2\pi$-periodicity 
\begin{align}
\tilde B^{2\pi,\mu}_n
= 
- \sigma^\mu_n (-1)^{\frac{1-\sigma^z_{n-1}\sigma^z_n}{2}+\frac{1-\sigma^z_n\sigma^z_{n+1}}{2}}
\label{eq:2D_TRS_bdy_spin}
\end{align}
for $\mu=x,y$. 
Importantly, the pumped spin operators $\tilde B^{\theta,\mu}_n$ depend only on the boundary spins, meaning that the $2\pi$-periodicity of $U^{\rm open}_\theta$ breaks only on the boundary. 

The pumped boundary spin operators (\ref{eq:2D_TRS_bdy_spin}) is also given by the local unitary on the boundary  
\begin{align}
U_{\rm bdy}=\prod_{n \in {\rm bdy}} e^{\frac{i\pi}{2}\frac{1-\sigma^z_n\sigma^z_{n+1}}{2}}. 
\end{align}
Thus, we have the operator relation 
\begin{align}
U_{2\pi}^{\rm open} = U_{\rm bdy}
\end{align}
up to a constant factor. 
$U_{\rm bdy}$ is known as the local unitary giving the Haldane chain for TRS $(\prod_n \sigma^x_n) {\cal K}$~\cite{CGLW13}.
Therefore, we can say the local unitary $U^{\rm open}_{\theta}$ pumps a Haldane chain phase by a period. 

We note that such a picture of the SPT phase pumped on the boundary for higher-dimensions is well known in the context of Floquet SPTs.~\cite{PM16,RH17,TV21}

\section{Adiabatic cycle in any dimension}
\label{sec:any_dim}
We integrate the results obtained in the previous sections and discuss a general theory of a kind of solvable model for any dimension. 
This section has much overlap with Ref.~\cite{RH17}, where the group cohomology construction of Floquet SPT drives in any dimension is given. 
The local unitary $U_\theta$ obtained in Sec.~\ref{eq:any_dim_local_unitary} is the same one in Ref.~\cite{RH17}. 

\subsection{Topological invariant from group cocycle}
Let $\Ket{\Psi_\theta}$ be an adiabatic cycle of gapped $G$-symmetric non-chiral ground state in $d$-spatial dimensions. 
Suppose that we have the inhomogeneous $(d+1)$-cocycle $\omega_\theta \in Z^{d+1}(G,U(1)_s)$ associated with the ground states $\Ket{\Psi_\theta}$. 
We also assume the $2\pi$-periodicity of $\omega_\theta$. 
Define the set of $\Z$ invariants from $\omega_\theta$ by 
\begin{align}
n(g_1,\dots,g_{d+1})
:= \frac{1}{2\pi i} \oint d \omega_\theta(g_1,\dots,g_{d+1}). 
\end{align}
The cocycle condition of $\omega_\theta$ implies that $n$ is a $(d+1)$-cocycle with the $\Z$-coefficient, i.e., $n \in Z^{d+1}(G,\Z_s)$. 
The $(d+1)$-cocycle $\omega_\theta$ is not unique. 
For a $2\pi$-periodic $d$-cochain $\alpha_\theta \in C^d(G,U(1)_s)$, $\omega_\theta$ and $\omega_\theta d\alpha_\theta$ represents physically the same ground states $\Ket{\Psi_\theta}$. 
Let 
\begin{align}
m(g_1,\dots,g_d)
:= 
\frac{1}{2\pi i}\oint d \alpha_\theta(g_1,\dots,g_d)
\end{align}
be the set of $\Z$ invariants of $\alpha_\theta$s.
The equivalence $\omega_\theta \sim \omega_\theta d\alpha_\theta$ means that the equivalence relation $n \sim n+dm$ by the $\Z$-valued $(d+1)$-coboundary $dm \in B^{d+1}(G,\Z_s)$. 
Therefore, given a cycle of $(d+1)$-cocycle $\omega_\theta$, one can define the set of integer invariants $[n]$ living in the group cohomology 
\begin{align}
H^{d+1}(G,\Z_s) = Z^{d+1}(G,\Z_s)/B^{d+1}(G,\Z_s).
\end{align}

\subsection{Group cohomology construction}
\label{eq:any_dim_local_unitary}
From the isomorphism $H^d(G,U(1)_s) \cong H^{d+1}(G,\Z_s)$, the invariant $[n]$ may be interpreted as the pump of an SPT phase in $(d-1)$-spatial dimensions. 
The isomorphism $H^d(G,U(1)_s) \to H^{d+1}(G,\Z_s)$ is given by the Bockstein homomorphism associated with the short exact sequence of the coefficients $\Z\to\R\to U(1)$. 
As we will see in this section, the Bockstein homomorphism gives us an exactly solvable lattice model of adiabatic cycles in the basis of Chen--Gu--Liu--Wen's construction.~\cite{CGLW13}

In this section we employ the homogeneous cochain $\nu \in C^d(G,U(1)_s)$. 
The relation to the inhomogeneous cochain $\omega$ is 
\begin{align}
\nu(g_0,g_1,\dots,g_d)
=\omega(g_0^{-1}g_1,\dots,g_{d-1}^{-1}g_d)^{s_{g_0}}.
\end{align}

Let $\nu(g_0,\dots,g_d) \in Z^d(G,U(1)_s)$ be an homogeneous $d$-cocycle, of which the equivalence class $[\nu]\in H^d(G,U(1)_s)$ is what we want to pump in $(d-1)$D. 
Let us denote $\nu(g_0,\dots,g_d)=e^{i\phi_\nu(g_0,\dots,g_d)}$ and introduce a lift 
\begin{align}
\phi_\nu(g_0,\dots,g_d) \to \tilde \phi_\nu(g_0,\dots,g_d) \in \R. 
\end{align}
The cocycle condition of $\nu$ ensures that the differential of $\tilde \phi_\nu$ is a $(d+1)$-cocycle of the $\Z$-coefficient $\frac{1}{2\pi}d\tilde \phi_\nu \in Z^{d+1}(G,\Z_s)$, and the equivalence class $[\frac{1}{2\pi}d\tilde \phi_\nu]$ gives the isomorphism $H^d(G,U(1)_s) \cong H^{d+1}(G,\Z_s)$. 

For an adiabatic cycle in $d$D, we introduce a $2\pi$-periodic homogeneous $(d+1)$-cocycle 
\begin{align}
\nu_\theta^{(d+1)}(g_0,\dots,g_{d+1})
= e^{\frac{i\theta}{2\pi} (d\tilde \phi_\nu)(g_0,\dots,g_{d+1})}. 
\end{align}
According to the recipe by Chen--Gu--Liu--Wen~\cite{CGLW13}, we get a model of $d$-dimensional exactly solvable model parameterized by $\theta$. 
To be precise, we consider a $d$-dimensional manifold with a triangulation with a branching structure equipped with the local Hilbert space spanned by the group basis $\ket{g}$ for $g \in G$. 
The $G$ action is defined by $\hat g \ket{h} = \ket{gh}^{s_g}$ for each site. 
The local unitary $U_\theta$ sending the trivial tensor product state to a state with the group cocycle $\nu_\theta^{(d+1)}$ is given by~\cite{CGLW13} 
\begin{align}
\tilde U_\theta
=\sum_{\{g_j\}}
\prod_{\Delta^{d}} 
\nu^{(d+1)}_\theta(g_*,g_0,\dots,g_d)^{\Delta^d|}
\ket{\{g_j\}} \bra{\{g_j\}}, 
\end{align}
where the product $\prod_{\Delta^d}$ runs over all the $d$-simplices, $|\Delta^d| \in \{\pm 1\}$ represents the orientation of the $d$-simplex $\Delta^d$, and $g_* \in G$ is an arbitrary fixed group element. 
The choice of $g_*$ does not affect the local unitary in bulk. 
This local unitary $\tilde U_\theta$ is manifestly $2\pi$-periodic, but breaks $G$-symmetry on the boundary~\cite{CGLW13}. 

Instead, we introduce alternative form of the local unitary. Let us expand the differential of $d \tilde \phi$. 
\begin{align}
&d\tilde \phi_\nu(g_*,g_0,\dots,g_d) \nonumber \\
&=\tilde \phi_\nu(g_0,g_1,\dots,g_d)-\tilde\phi_\nu(g_*,g_1,\dots,g_d) \nonumber \\
&+ \cdots +(-1)^{d+1} \tilde \phi_\nu(g_*,g_0,g_1,\dots,g_{d-1}).
\end{align}
We realize that except for the first term in the right-hand-side, this is the $d$-coboundary $d \tilde \alpha$ of the $(d-1)$-cochain $\tilde \alpha(g_0,\dots,g_{d-1}):=\tilde \phi_\nu(g_*,g_0,\dots,g_{d-1})$. 
We have 
\begin{align}
&\tilde U_\theta =\sum_{\{g_j\}}
\prod_{\Delta^{d}}\nonumber \\
& 
e^{
\frac{i\theta}{2\pi} |\Delta^d| 
\big(
\tilde \phi_\nu(g_0,\dots,g_d)-(d\tilde \alpha)(g_0,\dots,g_d)
\big)
}
\ket{\{g_j\}} \bra{\{g_j\}}, 
\end{align}
The coboundary term $d\tilde \alpha$ is canceled out each other with adjacent $d$-simplices in bulk. 
Therefore, the local unitary 
\begin{align}
U_\theta
=
\sum_{\{g_j\}}\prod_{\Delta^{d}} 
e^{
\frac{i\theta}{2\pi} |\Delta^d| 
\tilde \phi_\nu(g_0,\dots,g_d)
}
\ket{\{g_j\}} \bra{\{g_j\}}, 
\label{eq:generic_local_unitary}
\end{align}
provides the same action on the degrees of freedom strictly interior of bulk as that of $\tilde U_\theta$. 
The local unitary $U_\theta$ is the same one as in Ref.~\cite{RH17}. 

Compared to $\tilde U_\theta$, the local unitary $U_\theta$ has no periodicity for $\theta$, but preserves $G$ symmetry even in the presence of the boundary 
\begin{align}
\hat g U_\theta \hat g^{-1}
=U_\theta. 
\end{align}
This is from the homogeneous condition $\tilde \phi_\nu(gg_0,\dots,gg_d)=s_g \tilde \phi_\nu(g_0,\dots,g_d)$.
More generally, {\it for any function $\theta(\Delta^d)$ from the set of $d$-simplices to $\R$, the space-dependent local unitary 
\begin{align}
U[\theta]
=
\sum_{\{g_j\}}\prod_{\Delta^{d}} 
e^{
\frac{i\theta(\Delta^d)}{2\pi} |\Delta^d| 
\tilde \phi_\nu(g_0,\dots,g_d)
}
\ket{\{g_j\}} \bra{\{g_j\}}, 
\label{eq:generic_local_unitary_2}
\end{align}
is $G$ symmetric
\begin{align}
\hat g U[\theta] \hat g^{-1}
=U[\theta]
\end{align}
even in the presence of boundary.} 

The adiabatic Hamiltonian $H_\theta$ is defined by the unitary transformation by $U_\theta$ (or $\tilde U_\theta$) on the trivial Hamiltonian $H_0$ as in 
\begin{align}
H_\theta = U_\theta H_0 U_\theta^{-1}. 
\label{eq:generic_adiabatic_hamiltonian}
\end{align}
Here, $H_0$ is defined by the sum of local projectors onto the disordered state
\begin{align}
&H_0 = -\sum_j \ket{\phi}_j\bra{\phi}_j, \\
&\ket{\phi}_j = \frac{1}{\sqrt{|G|}}\sum_{g \in G}\ket{g}_j. 
\end{align}

In the rest of this section, we examine the properties of the adiabatic cycle $H_\theta$ as well as the local unitary $U_\theta$. 

\subsection{SPT phase pumped on the boundary}
\begin{figure}[!]
\centering
\includegraphics[width=\linewidth, trim=0cm 0cm 0cm 0cm]{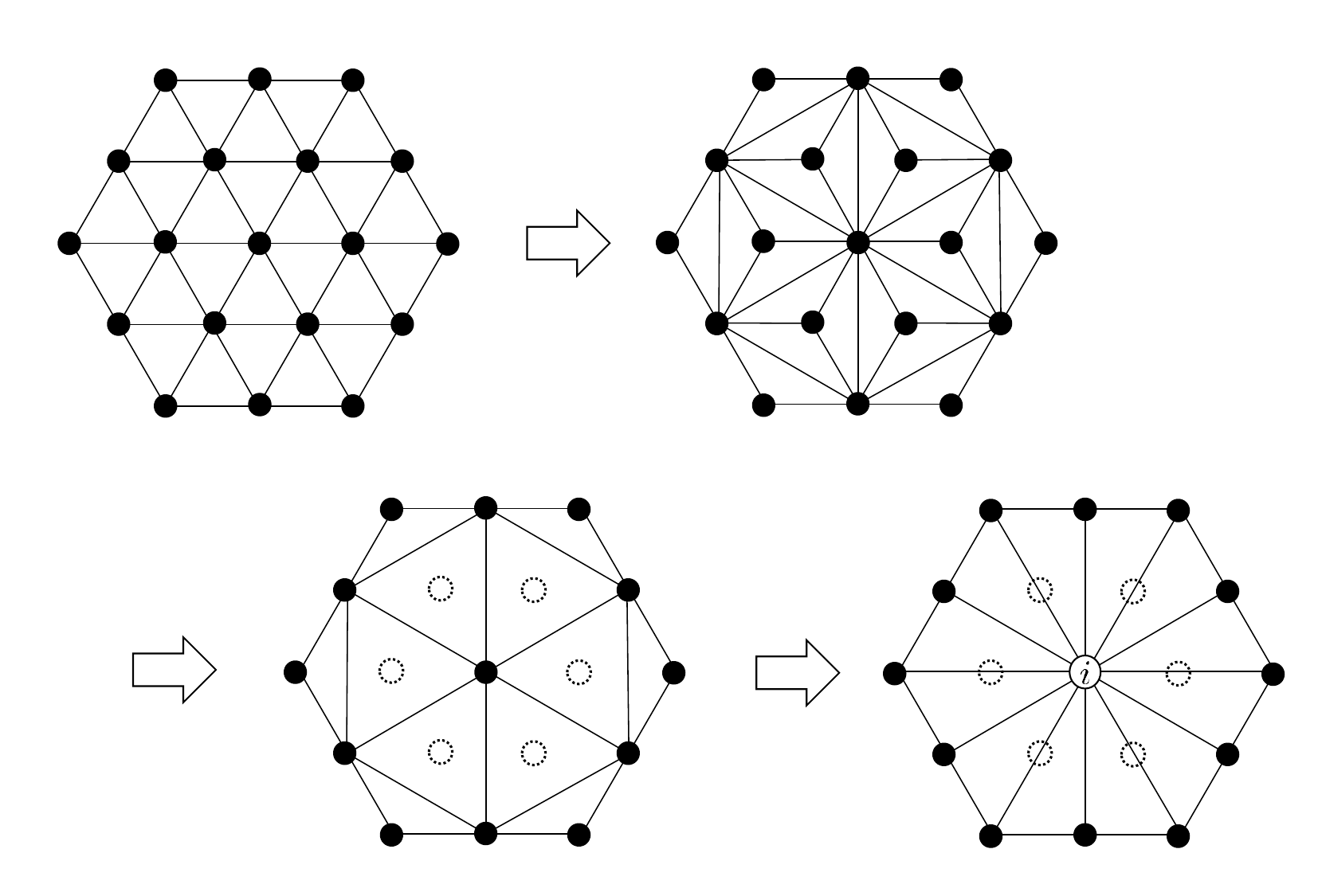}
\caption{The basic moves to remove internal sites. 
Here we omitted the branching structure.}
\label{fig:move}
\end{figure}

One can show that $U_\theta$ pumps the $(d-1)$D SPT phase with the $d$-cocycle $\nu$ directly~\cite{RH17}.
For a period $\theta=2\pi$, no ambiguity from the lift $\phi \to \tilde \phi$ remains, so we can safely write 
\begin{align}
U_{2\pi}
&= 
\sum_{\{g_j\}}\prod_{\Delta^d} 
e^{i |\Delta^d|
\phi_\nu(g_0,\dots,g_d)}
\ket{\{g_j\}} \bra{\{g_j\}} \nonumber\\
&= 
\sum_{\{g_j\}}\prod_{\Delta^d} 
\nu(g_0,\dots,g_d)^{|\Delta^d|}
\ket{\{g_j\}} \bra{\{g_j\}}.
\end{align}
Note that $U_{2\pi}$ is the identity for an closed space manifold because of the property $\prod_{\Delta^d} \nu(g_0,\dots,g_d)^{|\Delta^d|}=1$. 
With boundary, by using the basic moves (cocycle condition) to remove the internal sites except for a one site $i$, the amplitude is simplified as 
\begin{align}
\prod_{\Delta^d} 
\nu(g_0,\dots,g_d)^{|\Delta^d|}
=
\prod_{\Delta^{d-1}} 
\nu(g_i,g_0,\dots,g_{d-1})^{|\Delta^{d-1}|}, 
\label{eq:amp}
\end{align}
where the product $\prod_{\Delta^{d-1}}$ runs over the all boundary $(d-1)$-simplices. 
We illustrate the basic moves to remove the internal sites in Fig.~\ref{fig:move} for $d=2$. 

The amplitude (\ref{eq:amp}) is further simplified by using the cocycle condition 
\begin{align}
&\nu(g_i,g_0,\dots,g_{d-1})\nonumber\\
&=
\nu(g_*,g_0,\dots,g_{d-1})
\nu(g_i,g_*,g_1,\dots,g_{d-1})^{-1}\nonumber\\
&\cdots 
\nu(g_i,g_*,g_0,\dots,g_{d-2})^{(-1)^{d-1}}, 
\label{eq:amp_2}
\end{align}
where $g_* \in G$ is an arbitrary group element. 
In (\ref{eq:amp_2}), the factors including $g_i$ are canceled out with adjacent $d$-simplices, resulting in that $U_{2\pi}$ is the local unitary acting only on the boundary operators 
\begin{align}
&U_{2\pi}  = U_{\rm bdy}(\nu)\nonumber\\
&= 
\sum_{\{g_{n\in {\rm bdy}}\}}\prod_{\Delta^{d-1}} 
\nu(g_*,g_0,\dots,g_{d-1})^{|\Delta^{d-1}|}
\ket{\{g_n\}} \bra{\{g_n\}}, 
\label{eq:local_unitary_bdy}
\end{align}
where the sum $\sum_{g_n}$ runs over the boundary sites and the product $\prod_{\Delta^{d-1}}$ runs over the boundary simplices. 
$U_{2\pi}$ is nothing but the the local unitary giving the $(d-1)$D SPT phase labeled by the group $d$-cocycle $\nu \in Z^d(G,U(1)_s)$~\cite{CGLW13}.

\subsection{Texture induced SPT phase}
\label{sec:texture_SPT}
\begin{figure}[!]
\centering
\includegraphics[width=\linewidth, trim=0cm 0cm 0cm 0cm]{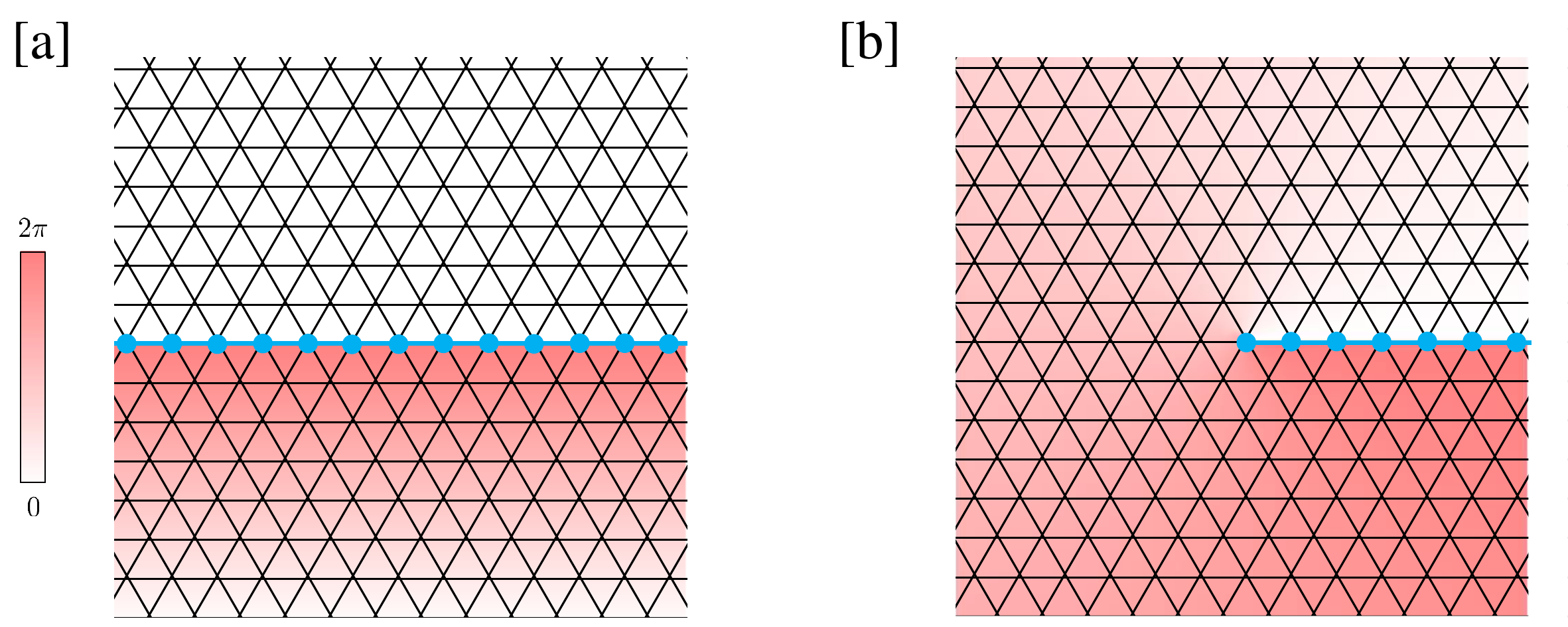}
\caption{
The intensity of the red color represents the function $\theta$. 
The codimension one surface $M_{d-1}$ is represented by the blue line in figures. 
}
\label{fig:texture}
\end{figure}

The local unitary $U_\theta$ can be used to generate a texture Hamiltonian which is exactly solvable. 
Let $\theta:\{\Delta^d\} \to [0,2\pi]$ be a function from $d$-simplices to the circle $[0,2\pi]$ where $2\pi$ and $0$ are identified.  
We consider the trial twist operator in the form 
\begin{align}
U[\theta]
= \sum_{\{g_j\}} \prod_{\Delta^d}
e^{\frac{i\theta(\Delta^d)}{2\pi} |\Delta^d| \tilde \phi_\nu(g_0,\dots,g_d)} \ket{\{g_j\}}\bra{\{g_j\}}. 
\end{align}
As seen in Sec.~\ref{sec:1D_texture}, because of the lack of the $2\pi$-periodicity of $U_\theta$ on the boundary, the twist operator $U[\theta]$ should be modified to give a smooth texture Hamiltonian. 
To do so, we introduce the $(d-1)$-dimensional manifold $M_{d-1}$ as the codimension one surface on which $\theta(\Delta^d)$ changes from $2\pi$ to 0, and insert the local unitary (\ref{eq:local_unitary_bdy}) over $M_{d-1}$. 
See Fig.~\ref{fig:texture} [a] for the illustration of $M_{d-1}$. 
Explicitly, 
\begin{align}
U(M_{d-1})
&=\sum_{\{g_{n \in M_{d-1}}\}} \prod_{\Delta^{d-1}\in M_{d-1}} \nonumber\\
&\nu(g_*,g_0,\dots,g_{d-1})^{|\Delta^{d-1}|} \ket{\{g_n\}}\bra{\{g_n\}}, 
\end{align}
where $n$ runs over $M_{d-1}$. 
The proper twist operator is defined as 
\begin{align}
U_{\rm twist}
= U(M_{d-1})^{-1} U[\theta]. 
\label{eq:local_unitary_texture}
\end{align}
Accordingly, the smooth texture Hamiltonian is given by 
\begin{align}
H_{\rm texture}
= U_{\rm twist} H_0 [U_{\rm twist}]^{-1}. 
\label{eq:texture_Ham}
\end{align}
The point is that we can explicitly write down the ground state 
\begin{align}
\Ket{\Psi_{\rm texture}}
=U_{\rm twist} \Ket{\Psi_0}, 
\end{align}
which we can concretely examine the properties of the texture ground state.

We note that our construction can also be applied for the vortex localized modes if $\theta(\Delta^d)$ has a singularity with nonzero winding number. 
For such cases, the codimension one surface $M_{d-1}$ has a boundary on the singularity of $\theta$. 
See Fig.~\ref{fig:texture} [b]. 
The local unitary sending the trivial Hamiltonian $H_0$ to the vortex Hamiltonian is defined in the same way as (\ref{eq:local_unitary_texture}).

We expect that the inserted local unitary $U(M_{d-1})$ is the origin of the emergence of the SPT phase in the texture. 
We consider this issue for 1D and higher dimensions separately. 

\subsubsection{1D}
We already see the emergence of $0D$ SPT phase for $G=\Z_2$ in Sec.~\ref{sec:1D_texture}. 
We here revisit this for the exactly solvable model (\ref{eq:generic_adiabatic_hamiltonian}). 

Given a 1-dimensional representation $e^{i \alpha_g} \in {\rm Hom}(G,U(1)_s) = Z^1(G,U(1)_s)$, we introduce a lift $\alpha_g \to \tilde \alpha_g \in \R$. 
The twist operator (\ref{eq:local_unitary_texture}) is given by 
\begin{align}
U_{\rm twist}
=
\sum_{\{g_j\}} e^{-i \alpha_{g_1}} \prod_{j=1}^N e^{\frac{i \theta_j}{2\pi} s_{g_j} \tilde \alpha_{g_j^{-1}g_{g+1}}} \ket{\{g_j\}}\bra{\{g_j\}}.
\end{align}
Here, $\theta_j$ is, for example, $\theta_j = \frac{2\pi j}{N}$. 
We confirm that the texture Hamiltonian $H_{\rm texture} = -\sum_j B_j = -\sum_j U_{\rm twist} P_j [U_{\rm twist}]^{-1}$ is indeed composed of smooth local terms, where $P_j = \frac{1}{|G|} \sum_{g,h} \ket{g}_j\bra{h}_j$, as follows. 
We have 
\begin{widetext}
\begin{align}
&B_j =\frac{1}{|G|}\sum_{g_{j-1},g_j,h_j,g_{j+1}}
e^{
\frac{i \theta_{j-1}}{2\pi}s_{g_{j-1}}
(\tilde \alpha_{g_{j-1}^{-1}g_j}
-\tilde \alpha_{g_{j-1}^{-1}h_j})
+\frac{i\theta_j}{2\pi}s_{g_j}
(\tilde\alpha_{g_j^{-1}g_{j+1}}
-\tilde\alpha_{h_j^{-1}g_{j+1}})
}
\ket{g_{j-1}g_jg_{j+1}}\bra{g_{j-1}h_jg_{j+1}}
\end{align}
for $j=2,\dots,N$, and 
\begin{align}
B_1 
&=\frac{1}{|G|}
\sum_{g_N,g_1,h_1,g_2}e^{-i(\alpha_{g_1}-\alpha_{h_1})}
e^{
\frac{i \theta_{N}}{2\pi}s_{g_N}
(\tilde \alpha_{g_{N}^{-1}g_1}
-\tilde \alpha_{g_{N}^{-1}h_1})
+\frac{i\theta_1}{2\pi}s_{g_1}
(\tilde\alpha_{g_1^{-1}g_{2}}
-\tilde\alpha_{h_1^{-1}g_{2}})
}\ket{g_{N}g_1g_{2}}\bra{g_{N}h_1g_{2}} \nonumber \\
&=\frac{1}{|G|} \sum_{g_N,g_1,h_1,g_2}
e^{
\frac{i\theta_1}{2\pi}
(\tilde\alpha_{g_1^{-1}g_{2}}
-\tilde\alpha_{h_1^{-1}g_{2}})
}
\ket{g_{N}g_1g_{2}}\bra{g_{N}h_1g_{2}}. 
\end{align}
Here we have used $\theta_{N}=2\pi$ and $e^{i \alpha_{gh}} = e^{i\alpha_g} e^{is_g\alpha_h}$. 
Note that without inserting the unitary $\sum_{g_1} e^{-i\alpha_{g_1}}\ket{g_1}\bra{g_1}$, the texture Hamiltonian is not smooth. 

The texture induced $0D$ state is evident from the symmetry property of the twist operator. 
We have 
\begin{align}
\hat g U_{\rm twist} \hat g^{-1}
=
\sum_{\{g_j\}} e^{-s_g i \alpha_{g_1}} \prod_{j=1}^N e^{s_g \frac{i \theta_j}{2\pi} s_{g_j}\tilde \alpha_{g_j^{-1}g_{g+1}}} \ket{\{gg_j\}}\bra{\{gg_j\}}
= e^{i\alpha_g} U_{\rm twist}.
\end{align}
This implies that the ground state $\Ket{\Psi_{\rm twist}} = U_{\rm twist} \Ket{\Psi_0}$ of the texture Hamiltonian $H_{\rm texture}$ has the $U(1)$ charge $e^{i\alpha_g}$ compared to the trivial ground state $\Ket{\Psi_0}$. 

\subsubsection{Higher dimensions}
To show that the texture Hamiltonian $H_{\rm texture}$ traps an SPT phase in one dimension lower, we explicitly compute the symmetry action on the boundary. 
Let $X_d$ be a $d$-dimensional space manifold with boundary and $M_{d-1}$ be the codimension one surface on which $\theta$ jumps from $2\pi$ to 0. 
Let us denote $g_{j \in \mathring{X_d}}$ and $g_{n \in \partial X_d}$ for group elements living inside bulk and boundary of $X_d$, respectively. 
The ground state manifold $\Ket{\Psi(\{g_{n\in \partial X_d}\})}$ of the texture Hamiltonian is explicitly written as 
\begin{align}
\Ket{\Psi(\{g_{n\in \partial X_d}\})}
=\sum_{\{g_{j\in \mathring{X_d}}\}}
\prod_{\Delta^{d-1} \in M_{d-1}} \nu(g_*,g_0,\dots,g_{d-1})^{-|\Delta^{d-1}|} 
\prod_{\Delta^d \in X_d} e^{\frac{i\theta(\Delta^d)}{2\pi} \tilde \phi_\nu(g_0,\dots,g_d)|\Delta^d|} \ket{\{g_j\},\{g_n\}}
\label{eq:GS_texture}
\end{align}
Note that the relative phases among the ground states $\Ket{\Psi(\{g_n\})}$ are arbitrary in general. 
We fix a set of relative phases as (\ref{eq:GS_texture}). 
Let us compute the symmetry action on the ground state manifold.
\begin{align}
\hat g\Ket{\Psi(\{g_{n\in \partial X_d}\})}
&=\sum_{\{g_{j\in \mathring{X_d}}\}}
\prod_{\Delta^{d-1} \in M_{d-1}} \nu(g_*,g_0,\dots,g_{d-1})^{-s_g|\Delta^{d-1}|} 
\prod_{\Delta^d \in X_d} e^{\frac{s_g i\theta(\Delta^d)}{2\pi} \tilde \phi_\nu(g_0,\dots,g_d)|\Delta^d|} \ket{\{gg_j\},\{gg_n\}}^{s_g} \nonumber\\
&=\sum_{\{g_{j\in \mathring{X_d}}\}}
\prod_{\Delta^{d-1} \in M_{d-1}} \nu(gg_*,\tilde g_0,\dots,\tilde g_{d-1})^{-|\Delta^{d-1}|} 
\prod_{\Delta^d \in X_d} e^{\frac{i\theta(\Delta^d)}{2\pi} \tilde \phi_\nu(\tilde g_0,\dots,\tilde g_d)|\Delta^d|} \ket{\{g_j\},\{gg_n\}}^{s_g}. 
\end{align}
Here, we used the homogeneous condition of $\nu$ and $\tilde \phi_\nu$ and introduced the notation 
\begin{align}
\tilde g_x =\left\{\begin{array}{ll}
g_x &(x \in \mathring X_d), \\
gg_x &(x \in \partial X_d). \\
\end{array} \right.
\end{align}
At this stage, we find that the symmetry acts only on the codimension one surface $M_{d-1}$, and thus, the problem is completely reduced to how the symmetry acts on the boundary of $M_{d-1}$, which is well-known. 
See, for example, Ref.~\cite{EN14}. 
For self-contentedness, we further compute the boundary symmetry action. 
Using the cocycle condition 
\begin{align}
&\nu(gg_*,\tilde g_0,\dots,\tilde g_{d-1})
\nu(g_*,\tilde g_0,\dots,\tilde g_{d-1})^{-1} \nu(g_*,gg_*,\tilde g_1,\dots,\tilde g_{d-1})
\nonumber\\
&\qquad \qquad 
\nu(g_*,gg_*,\tilde g_0,\tilde g_2,\dots,\tilde g_{d-1})^{-1}\cdots 
\nu(g_*,gg_*,\tilde g_0,\dots,\tilde g_{d-2})^{(-1)^{d+1}}=1, 
\end{align}
we have 
\begin{align}
\hat g\Ket{\Psi(\{g_{n\in \partial X_d}\})}
&=\prod_{\Delta^{d-2} \in \partial M_{d-1}} 
\nu(g_*,gg_*,gg_0,\dots,gg_{d-2})^{|\Delta^{d-2}|}\nonumber \\
&\sum_{\{g_{j\in \mathring{X_d}}\}}
\prod_{\Delta^{d-1} \in M_{d-1}} \nu(g_*,\tilde g_0,\dots,\tilde g_{d-1})^{-|\Delta^{d-1}|} 
\prod_{\Delta^d \in X_d} e^{\frac{i\theta(\Delta^d)}{2\pi} \tilde \phi_\nu(\tilde g_0,\dots,\tilde g_d)|\Delta^d|} \ket{\{g_j\},\{gg_n\}}^{s_g}\nonumber \\
&= {\cal N}_{\partial M_{d-1}}(g) {\cal S}_{\partial X_d}(g) {\cal K}^{s_g} \Ket{\Psi(\{g_n\}}.
\end{align}
Here, we have introduced the local unitaries ${\cal N}_{\partial M_{d-1}}$ and ${\cal S}_{\partial X_d}$ acting on the ground state manifold $\Ket{\Psi(\{g_n\})}$ which have supports on $\partial M_{d-1}$ and $\partial X_d$, respectively, by~\cite{EN14}
\begin{align}
&{\cal S}_{\partial X_d}(g) \Ket{\Psi(\{g_n\}}
= \Ket{\Psi(\{gg_n\}}, \\
&{\cal N}_{\partial M_{d-1}}(g) \Ket{\Psi(\{g_n\}}
= \prod_{\Delta^{d-2} \in \partial M_{d-1}} 
\nu(g_*,gg_*,g_0,\dots,g_{d-2})^{|\Delta^{d-2}|}
\Ket{\Psi(\{g_n\}}.
\end{align}
The local unitary ${\cal N}_{\partial M_{d-1}}(g) {\cal S}_{\partial X_d}(g) {\cal K}^{s_g}$ (restricted to $\partial M_{d-1}$) is known as an anomalous symmetry action of the $(d-1)$D SPT phase with the cocycle $\nu \in Z^{d}(G,U(1)_s)$.
Thus, we have shown that the texture Hamiltonian (\ref{eq:texture_Ham}) indeed traps the $(d-1)$D SPT phase. 
\end{widetext}

\subsection{Examples of local unitary}
We illustrate the local unitary  (\ref{eq:generic_local_unitary_2}) with a few examples. 
See also Ref.~\cite{RH17}. 

\subsubsection{1D, $\Z_2$ symmetry}
Let us consider the unitary symmetry group  $G=\Z_2=\{e,\sigma\}$. 
There is only one nontrivial representation of $\Z_2$, $e^{i\alpha_\sigma}=-1$. 
A lift is given by $\tilde \alpha_\sigma=\pi$. 
The local unitary (\ref{eq:generic_local_unitary}) reads 
\begin{align}
U_\theta
=
\sum_{\{\sigma_j\}}
\prod_{j}
e^{\frac{i\theta}{2} \frac{1-\sigma_j\sigma_{j+1}}{2}}
\ket{\{\sigma_j\}}\bra{\{\sigma_j\}}.
\end{align}
This is nothing but the local unitary (\ref{eq:U_bulk}) discussed in Sec.~\ref{sec:1DZ2}. 

\subsubsection{2D, $\Z_2^T$ symmetry}
Let us consider $\Z_2^T$ time-reversal symmetry. 
The inhomogeneous cocycle $\omega$ representing the nontrivial group cohomology $H^2(\Z_2^T,U(1)_s)=\Z_2$ is 
\begin{align}
\omega(g,h)
=\left\{\begin{array}{ll}
-1 & (g=h=\sigma), \\
1 & ({\rm else}). 
\end{array}\right.
\end{align}
Accordingly, a lift is given by 
\begin{align}
\tilde \phi(g,h)
=\left\{\begin{array}{ll}
\pi & (g=h=\sigma), \\
0 & ({\rm else}). 
\end{array}\right.
\end{align}
The local unitary (\ref{eq:generic_local_unitary}) is 
\begin{align}
U_\theta
&=
\sum_{\{\sigma_j\}}
\prod_{\Delta^2}
e^{
\frac{i\theta}{2} |\Delta^2| 
\frac{1-\sigma_0 \sigma_1}{2}
\frac{1-\sigma_1\sigma_2}{2}
}
\ket{\{\sigma_j\}}\bra{\{\sigma_j\}} \nonumber\\
&=
\prod_{\Delta^2}
e^{
\frac{i\theta}{2} |\Delta^2| 
\frac{1-\sigma^z_0 \sigma^z_1}{2}
\frac{1-\sigma^z_1\sigma^z_2}{2}
}.
\end{align}
This differs from the local unitary (\ref{eq:2D_TRS_local_unitary}) discussed in Sec.~\ref{sec:2D_model}, but supposed to belong to the same adiabatic cycle.

\subsubsection{3D, $\Z_2$ symmetry}
We here present only one example of adiabatic cycle in 3D that pumps a nontrivial 2D SPT phase on the boundary. 
For $G=\Z_2$, SPT phases are classified by $H^3(\Z_2,U(1)) = \Z_2$ and a representative inhomogeneous three-cocycle $\omega \in Z^3(\Z_2,U(1))$ is given by 
\begin{align}
\omega(g,h,k)=\left\{\begin{array}{ll}
-1 & (g=h=k=\s), \\
1 & ({\rm else}). \\
\end{array}\right.
\end{align}
Correspondingly, the local unitary in 3D is given by 
\begin{align}
U_\theta
=
\prod_{\Delta^3}
e^{\frac{i\theta}{2} |\Delta^3| \frac{1-\sigma^z_0 \sigma^z_1}{2}\frac{1-\sigma^z_1\sigma^z_2}{2}\frac{1-\sigma^z_2\sigma^z_3}{2}}. 
\end{align}

\section{Summary}
\label{sec:sum}

We studied adiabatic cycles in quantum spin systems with unique gapped ground states. 
Through the detailed calculation of the toy models in one and two dimensions and the MPS representation for one dimension, we show that the set of winding numbers of $(d+1)$-cocycle in  $Z^{d+1}(G,U(1)_s)$, which characterizes a unique gapped ground state with $G$ symmetry, serves as topological invariants of adiabatic cycles. 
These topological invariants are found to live in the group cohomology $H^{d+1}(G,\Z_s)$.
The Bockstein homomorphism $H^d(G,U(1)_s) \to H^{d+1}(G,\Z_s)$ gives us an exactly solvable model of the adiabatic cycle by Chen-Gu-Liu-Wen's group cohomology construction~\cite{CGLW13}. 
The obtained one-parameter local unitary is the same one as that in Ref.~\cite{RH17}. 
We demonstrated that an SPT phase emerges at the spatial texture on which the adiabatic parameter winds a period.






{\it Note added.}
After the first version of this paper was submitted to the arXiv, the author was informed of the video~\cite{Her}. 
Some points in \cite{Her} overlap with Sec.~\ref{sec:1DZ2} in this paper.

{\it Acknowledgment.}
We thank Takahiro Morimoto and Ryohei Kobayashi for useful discussions. 
We also thank Shuhei Ohyama and Masatoshi Sato for helpful discussions with the related collaboration. 
K.S. is supported by JST CREST Grant No. JPMJCR19T2 and JSTPRESTO  Grant  No.  JPMJPR18L4. 


\bibliography{refs}

\end{document}